\pdfoutput=1

\documentclass{article}
\usepackage[utf8]{inputenc}
\usepackage[super,sort&compress,comma]{natbib} 
\usepackage{graphicx, wrapfig, lipsum}
\usepackage{authblk}
\usepackage[format=plain,justification=justified,singlelinecheck=false,labelfont=bf,labelsep=space]{caption}
\usepackage{subcaption}
\usepackage{float}
\usepackage{xurl}
\usepackage{amsmath}
\usepackage{amsfonts}
\usepackage{dirtree}
\usepackage{pdflscape}
\usepackage{geometry}

\renewenvironment{abstract}
{\begin{quote}
\noindent \rule{\linewidth}{.5pt}\par{}}
{\smallskip\noindent \rule{\linewidth}{.5pt}
\end{quote}
}

\title{AlabOS: A Python-based Reconfigurable Workflow Management Framework for Autonomous Laboratories}

\author[1,2,5]{Yuxing Fei}
\author[1,2,5]{Bernardus Rendy}
\author[3,5]{Rishi Kumar}
\author[3]{Olympia Dartsi}
\author[1,3]{Hrushikesh P. Sahasrabuddhe}
\author[2]{Matthew J. McDermott}
\author[1,2]{Zheren Wang}
\author[1,2]{Nathan J. Szymanski}
\author[2]{Lauren N. Walters}
\author[2]{David Milsted}
\author[2,\textdagger]{Yan Zeng}
\author[3,\textdagger]{Anubhav Jain}
\author[1,2,\textdagger]{Gerbrand Ceder}
\affil[1]{Department of Materials Science \& Engineering, University of California, Berkeley, CA 94720, USA}
\affil[2]{Materials Sciences Division, Lawrence Berkeley National Laboratory, Berkeley, CA 94720, USA}
\affil[3]{Energy Technologies Area, Lawrence Berkeley National Laboratory, Berkeley, CA 94720, USA}
\affil[5]{Equal contribution}
\affil[$\dagger$]{Corresponding author: yanzeng@lbl.gov, ajain@lbl.gov and gceder@berkeley.edu}

\date{}

\begin{document}
\maketitle
\newpage
\begin{abstract}
The recent advent of autonomous laboratories, coupled with algorithms for high-throughput screening and active learning, promises to accelerate materials discovery and innovation. As these autonomous systems grow in complexity, the demand for robust and efficient workflow management software becomes increasingly critical. In this paper, we introduce AlabOS, a general-purpose software framework for orchestrating experiments and managing resources, with an emphasis on automated laboratories for materials synthesis and characterization. AlabOS features a reconfigurable experiment workflow model and a resource reservation mechanism, enabling the simultaneous execution of varied workflows composed of modular tasks while eliminating conflicts between tasks. To showcase its capability, we demonstrate the implementation of AlabOS in a prototype autonomous materials laboratory, A-Lab, with around 3,500 samples synthesized over 1.5 years.

\end{abstract}
\newpage
\section{Introduction}
Computational and data-driven approaches have shaped a new paradigm for materials discovery, leveraging recent advances in high-performance computing and machine learning. \cite{jain2013commentary, schmidt2019recent, choudhary2022recent, butler2018machine, schleder2019dft, morgan2020opportunities, kirklin2015open} Despite these advances, the experimental realization of computationally predicted compounds remains a slow, yet essential step in materials discovery and development\cite{SumpterCNPGM, ChamorroACR, WangNatSynth}. Substantial progress has been made to increase experimental throughput while reducing human involvement, as demonstrated by several automated laboratories that utilize synthesis methods based on flow chemistry \cite{steiner2019organic, hartrampf2020synthesis, manzano2022autonomous, bennett2024autonomous}, thin-film deposition \cite{macleod2020self, kusne2020fly}, solution-based synthesis \cite{burger2020mobile}, and solid-state synthesis \cite{Szymanski2023, chen2024navigating}. These laboratories have also been integrated with predictive computational tools such as high-throughput density functional theory (DFT) calculations \cite{correa2018accelerating, saal2013materials, jain2013commentary, liu2021application}, machine learning algorithms for automated interpretation of characterization data \cite{szymanski2021probabilistic, stanev2018unsupervised, chen2021automating, liu2017deep, oviedo2019fast}, and optimization algorithms that can perform decision making \cite{szymanski2023autonomous, aykol2021rational, strieth2023delocalized, granda2018controlling, xie2022toward}. The integration of computational modeling and artificial intelligence into autonomous laboratories gives rise to what is known as self-driving laboratories (SDLs), which can explore material spaces with minimal human intervention\cite{abolhasani2023rise, martin2023perspectives, xie2023toward}. The rising prominence of SDLs has elevated the significance of robotics and machine learning for materials research. To accommodate this changing landscape from the human-centric experimentation diagram, improved tools are needed to manage robotics and the data they rapidly produce. \cite{raccuglia2016machine, zimmerman2014data, talirz2020materials}\\

In designing management software, one can learn from the automated workflows developed for high-throughput DFT calculations \cite{jain2015fireworks, himanen2019data, curtarolo2012aflow}. These workflows provide a high-level user interface from which materials and calculation types can be specified, abstracting the lower-level tasks required to perform such calculations \cite{mathew2017atomate, huber2021common}. These computational management software need to effectively organize large datasets, a task similarly required for experimental management. However, orchestrating an autonomous laboratory creates additional challenges, in particular, requiring the seamless integration of tasks and data with physical hardware and experimental samples, which is crucial for maintaining uninterrupted operation. \\

Several workflow management programs have previously been developed and implemented in different autonomous laboratories with diverse applications. For example, ChemOS 2.0 \cite{sim2023chemos} proposes an integral platform between DFT calculation, Bayesian optimization, and automated equipment communicated through SiLA2 protocol \cite{Juchli2022}. Helao \cite{rahmanian2022enabling} and Helao-async \cite{guevarra2023orchestrating} present a highly modularized design by implementing each component as a web server, enabling workflow management across laboratories. Bluesky \cite{allan2019bluesky} has been applied to synchrotron facilities, with a strong emphasis on controlling synchrotron characterization hardware, to improve data collection efficiency but not targeted to materials synthesis. Additional software packages have supported laboratories involving automated experiments, featured with expandability and reconfigurability across different experiment stations \cite{fakhruldeen2022archemist}, standardized data specification and storage \cite{pendleton2019experiment}, and seamless integration with other science tools as a ``science factory'' \cite{vescovi2023towards}. These software solutions have successfully orchestrated experiment workflows in autonomous laboratories within their domains. \\

On the other hand, recent trends toward increased throughput and greater complexity in autonomous laboratories necessitate workflow software to manage more diverse workflows with a higher volume of autonomous devices. For instance, in our use case as well as others aiming to boost overall throughput, a laboratory can feature multiple equivalent devices for a single task, such as several furnaces for heating or two X-ray diffractometers for characterization. Due to the lack of knowledge about the availability of devices during runtime, the workflow management system must dynamically allocate the devices to each task based on their needs, thereby maximizing the utilization rate of the autonomous laboratory. Moreover, as general-purpose research laboratories often run different workflows with varied task sequences simultaneously, a reconfigurable workflow model has become essential, with certain parts of the workflow being executed in parallel. For example, after a sample is synthesized, multiple characterizations can be performed concurrently while storage operation has to wait until all the characterizations are done. Depending on the needs of the experiments, different experiment procedures need to be executed in the same laboratory setting. Thus, an expressive workflow model is required to effectively encode such task order dependencies while preserving flexibility when composing the workflow. Efforts have been made to handle some of these challenges. For example, a scheduling strategy is developed to tackle the multi-device problem, which requires an accurate time model for each task to ensure efficiency.\cite{zhou2024multi} However, only by addressing all of these practical needs, a workflow management software can fit better into an autonomous laboratory with increased complexity and throughput and allow researchers greater freedom to push the boundaries of autonomous experimentation.\\

Herein, we present AlabOS, a versatile and accessible workflow management framework for autonomous laboratories. The system features a graph-based experimental workflow model with tasks being the nodes and task dependencies being the edge. A central resource manager is built into the system, to track the status of devices and allocate resources properly to each task, thus eliminating possible conflicts between tasks running at the same time in the laboratory. Furthermore, AlabOS proposes the concept of sample position, representing a position in the laboratory that can hold one sample. By tracking the sample's position, AlabOS makes it possible to track individual samples in real time. This Python-based software is platform-independent and designed to be user-friendly, requiring basic knowledge about database and parallel programming for community adoption. Considering the errors and maintenance demands in the daily operations of an autonomous laboratory, a status monitoring and notification system is built into the software along with a browser-based graphic user interface (GUI). AlabOS serves as a general framework for managing workflows, designed to simplify the programming required to establish an autonomous laboratory. At the same time, it maintains the flexibility to accommodate various workflows within the laboratory. Other features of AlabOS include:

\begin{itemize}
    \item Provide a solution for autonomous laboratory workflow management, with common functionalities like resource management, and notification built in.
    \item Built with MongoDB (NoSQL) backend, supporting a flexible schema, and allowing standardized continuous development of each task input, output, and data storage operations.
    \item A standard way to define devices and tasks with base classes when setting up a new lab with AlabOS. A simulation mode is also built in, allowing quick debugging of the devices and tasks.
    \item Submission and status monitoring APIs with JSON format. The users can set up scripts for job submission and queries on top of AlabOS. The input and output of the experiments can be validated by Pydantic \cite{pydantic2024} before being stored in the database.
\end{itemize}

The AlabOS system is actively deployed in the A-Lab \cite{Szymanski2023}, an autonomous laboratory for inorganic materials synthesis housed at Lawrence Berkeley National Laboratory (LBNL). At the time of writing, it has synthesized and characterized over 3,500 distinct samples under the control of AlabOS.

\section{Key Concepts and System Design}
\subsection{Status representations}
\begin{figure}[htp]
    \centering
    \includegraphics[width=0.6\linewidth]{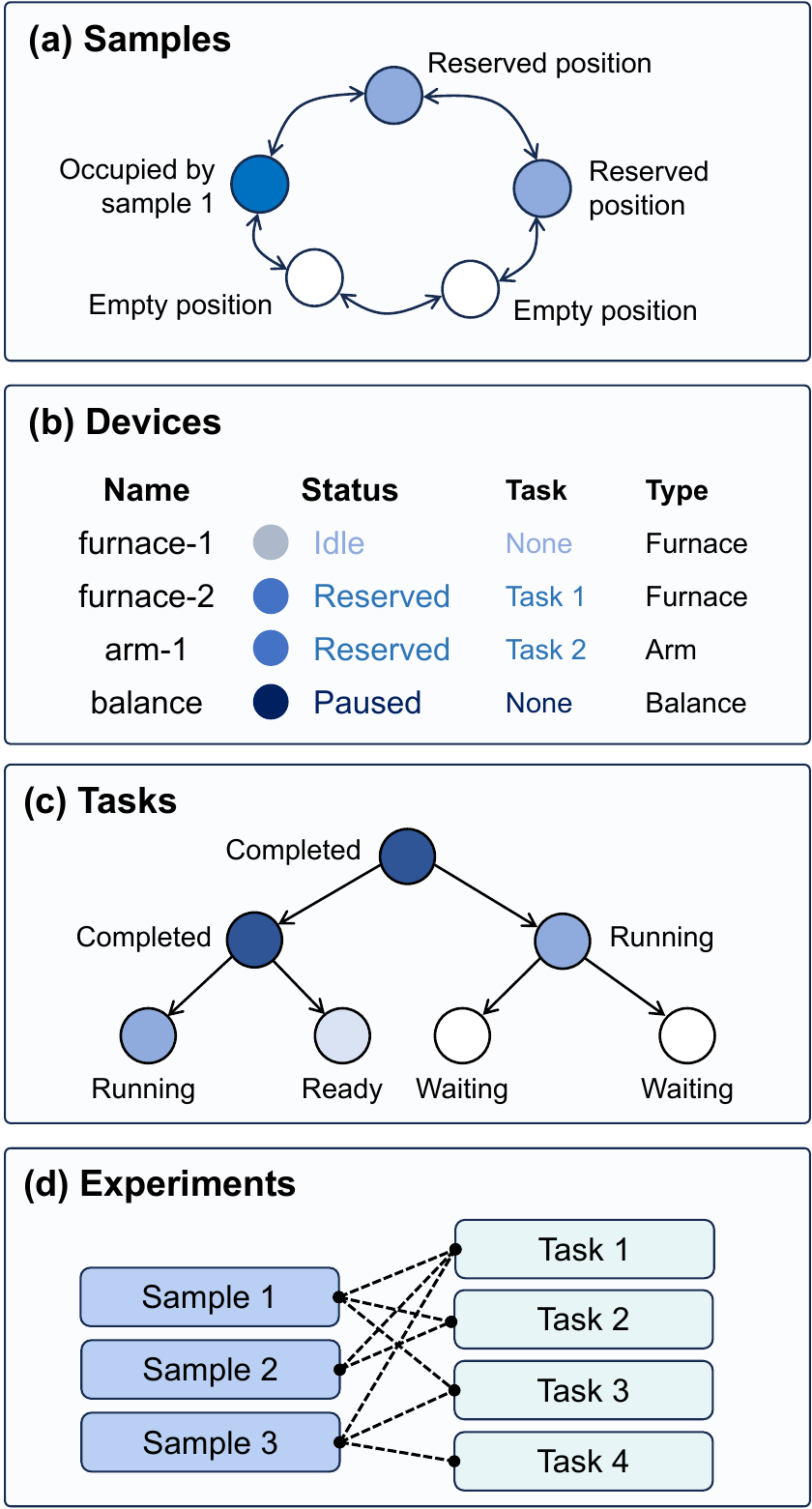}
    \caption{Schematic representation of lab states managed within the AlabOS system. The lab status is divided into four collections: (a) The sample collection tracks the sample’s position and identifies the task reserving that position; (b) the device collection records the status of the device and the task assigned to it; (c) the task collection records the order of execution for different experimental tasks. The order of tasks is encoded in a graph data structure to ensure sequential execution is performed in the correct order; (d) the experiment collection records a batch of samples and their associated tasks.}
    \label{fig:concept}
\end{figure}

In AlabOS, a laboratory is represented as a combination of samples, devices, tasks, and experiments entities (Fig.~\ref{fig:concept}). Each entity record is maintained in a separate collection in a MongoDB instance and updated during runtime to synchronize with the autonomous laboratory.\\

A sample entity describes its name, position, and other metadata specified during submission (e.g., the composition of the sample and the project that the sample belongs to).  A sample position is a space in the laboratory that can be occupied by one sample at a time. The sample positions are defined by operators before the system is up. When robots (or humans) move a sample, the positions of the samples in the database are updated to track its physical location throughout the experiments. Every sample is assigned a human-readable name and a unique global ID for tracking purposes. Within autonomous laboratories, transferring a sample from one container to another is common. In this case, the old container will be disposed of in the physical laboratory while the sample itself will be updated to the position of the new container. When a sample is completely removed from the laboratory, its position will be set to null while the database entry describing its position history is kept for future reference.\\

A device refers to a piece of hardware to process or collect data from sample(s). Each device entity is linked to hardware in the lab that can send commands, perform physical operations, and collect data. Some examples of these operations in materials synthesis include sample dispensing, weighing, heating, and grinding.
In AlabOS, a device is defined in Python code that provides the methods to communicate with the data transfer protocols like MODBUS \cite{swales1999open}, TCP/IP, and serial. Each device will have an entry in the database that indicates its status: whether it is occupied by a task, or whether it is paused for maintenance.\\

A task entity contains the procedure to execute a sequence of operations on a set of samples using the specified device.
In a task, all relevant devices are orchestrated to achieve a high-level objective. For instance, to carry out a heating task, the process begins by sending a request to the furnace to open its door. Following this, a robotic arm moves the designated samples into the now-open furnace. After the samples are loaded, a program is activated to heat the furnace to a temperature specified by the operator. Considering the parallel nature of an autonomous laboratory, where multiple samples may be processed simultaneously, a resource assignment mechanism is used to avoid conflicts between tasks in resource assignment. Before initiating control over one device or sample, each task has to request devices and sample positions. Once the resources are assigned, the task runs the procedures. After completion, the associated resources are released, allowing other tasks to request and use them. Apart from processing, a task could also be used for data generation and analysis for characterization or decision-making. In this case, the output of one task is directed as an input for another task by utilizing sample metadata as an information proxy.\\

An experiment is composed of one or multiple series of tasks to obtain conclusions. When submitting an experiment, the operator can specify a directed acyclic graph (DAG) of tasks to be performed on each sample. The execution sequence is guaranteed by the directed edge in DAG, where each node represents a task and each edge represents its order. A task can only be started when it does not have any unfinished parent tasks, which are defined to happen before this task. Relationships among tasks are stored in the task collection, where each task has a previous\_tasks field and a next\_tasks field. Whenever a task is completed, the system will initiate, request resources, and launch any descendent tasks when they are ready.

To maximize throughput, an autonomous laboratory often processes samples in batches. In AlabOS, an experiment can contain multiple samples and tasks. A task can accept one or more samples as input, depending on its predefined capacity. For example, if a furnace has a capacity for up to eight samples at a time, the heating task will not accept more than eight samples. \\

\subsection{AlabOS architecture}
\begin{figure}[htp]
    \centering
    \includegraphics[width=\linewidth]{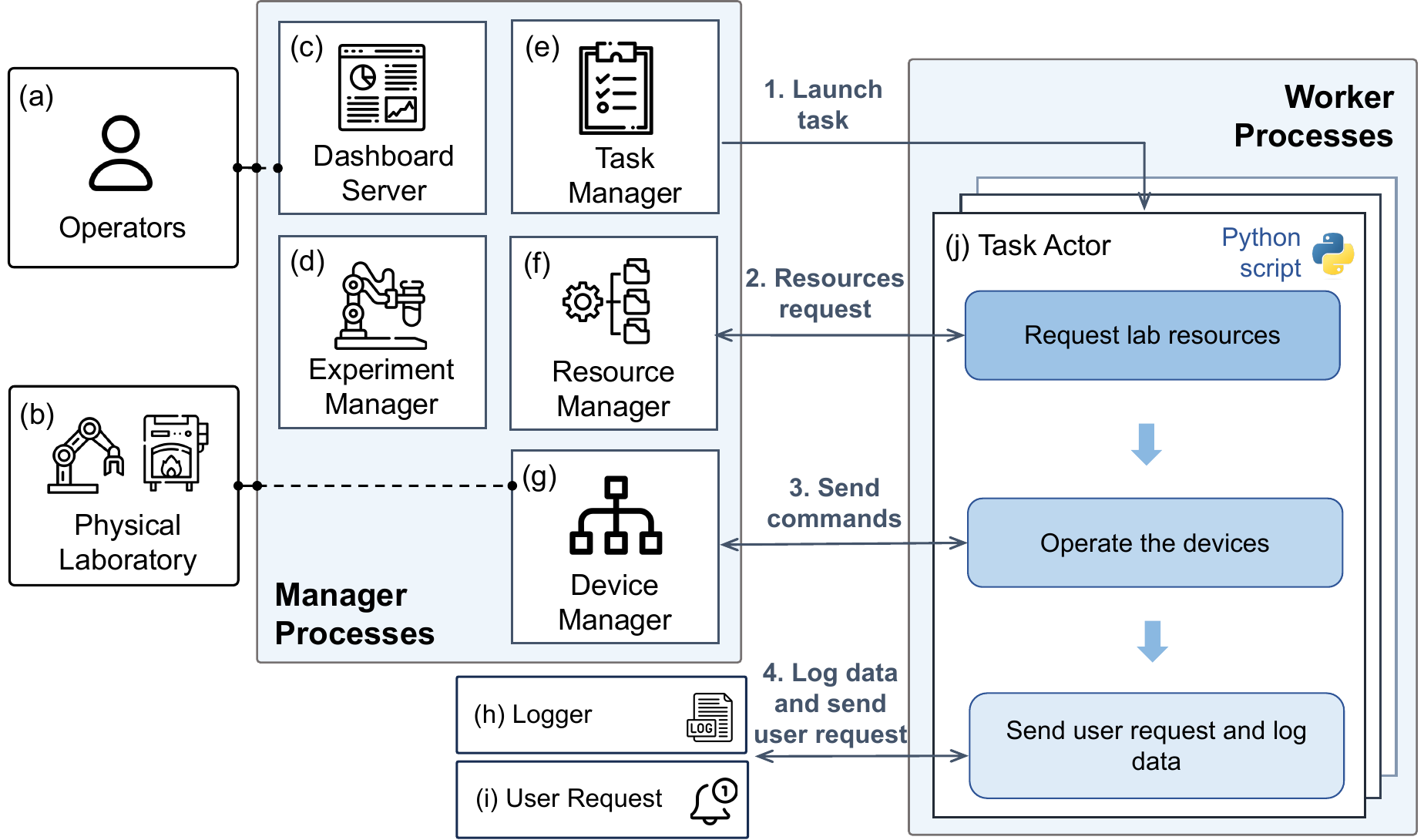}
    \caption{An architecture diagram for AlabOS. (a) Human operators submit experiments, monitor the laboratory state, and manage user requests. (b) The physical laboratory performs these tasks via manager processes. These include (c) the dashboard server that hosts the user interface and APIs for monitoring and controlling the experiment workflows; (d) the experiment manager that transforms high-level requests into specific tasks; (e) the task manager that launches each task and monitors its status; (f) the resource manager that assigns and tracks the status of available lab devices and sample positions; (g) the device manager that directs traffic between task actors and physical lab devices. Outside the manager processes, a logger module (h) logs the runtime information and saves task results into a central database; a user request module (i) requests user interventions and sends user responses back to the requester. In the worker processes, each task is carried out within a task actor (j) that can request resources, send commands to devices, log data, and initiate user requests.}
    \label{fig:fig1}
\end{figure}

The AlabOS is designed to be a manager-worker architecture (Fig.~\ref{fig:fig1}).
When a task is ready to run within this architecture, a worker object (named task actor) is instantiated in a new process by a task manager that executes the task logic. The managers monitor the status of the task actors and respond to each actor's requests for resources in the laboratory. Since the task actors only communicate with the managers, the interaction between ongoing laboratory tasks is eliminated. As a result, no conflicts between concurrent tasks (race conditions) need to be considered when defining the tasks' logic. The core services also run a dashboard server, which provides operators with a browser-based GUI and a set of APIs. The lab operator can easily submit, view, and cancel experiments in the autonomous laboratory using these interfaces.\\

There are four manager processes used to monitor and manipulate the status of the laboratory at different levels: experiments, tasks, devices, and resources (the assignment of sample positions and devices). In addition, a dashboard server is working as the manager process to receive commands from the operators. All the managers communicate with one another through a database instance that is hosted on either a local computer or a cloud-based server. The experiment manager (Fig.~\ref{fig:fig1}d) receives experimental submissions from the dashboard server (Fig.~\ref{fig:fig1}c) and parses them into task graphs. The task manager (Fig.~\ref{fig:fig1}e) verifies and launches these tasks ready in the laboratory. It also monitors the status of each task as it is carried out. The task manager is alerted when a task is completed, or an error is encountered. It then marks all subsequent tasks as being ready or canceled accordingly. The resource manager (Fig.~\ref{fig:fig1}f) responds to the tasks' requests to occupy certain devices and sample positions in the lab. When a request is received, the manager checks the availability of all requested devices and sample positions. If the request can be met, the resource manager assigns the devices and sample positions to the task. The assignment will be valid until the signal is received that the task's resources are released. The device manager (Fig.~\ref{fig:fig1}g) is the intermediate layer between each task and the physical lab device it affects. When a task requests the device manager to send a command to some device, it first checks whether the device is occupied before sending the commands required to complete the task. \\

To interface with the manager processes in a user-friendly way, a dashboard server (Fig.~\ref{fig:fig1}c) with a GUI is provided. This allows the operator to control and monitor the progress of any experiments running in the laboratory, while also showing the current states of all devices in the lab. In addition to the GUI, this dashboard provides an API that can receive new experiment submissions in a JSON format \cite{pezoa2016foundations}. Each experiment submission is validated using \textit{Pydantic} \cite{pydantic2024} models to ensure the correctness of all formats and values. \\

In the worker processes, many task actors (Fig.~\ref{fig:fig1}j) run simultaneously to execute different tasks in the laboratory. In AlabOS, each task is configured in advance before the start of a new experiment. All tasks are defined as Python objects inherited from the BaseTask class available in the AlabOS package, providing universal methods for interacting with manager processes, the logger, and the user request module. To execute lab operations, the task actor must first request to occupy some devices and sample positions from the resource manager. When the request is approved, it continues to send commands to devices via the device manager while also updating each sample's positions in the database. The data and device signal (e.g., the real-time temperature in a furnace) generated during the task will be logged to the database by the logger module (Fig.~\ref{fig:fig1}h). If the task requires any human intervention (for example, recovering a robot arm from error, or replacing consumables), it will generate a notification via the user request module, which the human operators can acknowledge once it is resolved.

\newpage

\section{Features and Implementations}
Several features are incorporated into AlabOS for resource management,  human-machine interaction (HMI), lab device control, and data organization to ensure the autonomous laboratory runs efficiently and smoothly. These are detailed in the next few sections. 


\subsection{Resource management}
Unlike traditional materials research laboratories, where human experts perform various tasks (like sample preparation, heating, and characterization) serially, autonomous laboratories usually take a more distributed and parallelized approach. For example, a robot may dispense and weigh precursor powders while an automated furnace is heating a separate batch of samples that have already been weighed. There is typically a queue of pending tasks that the laboratory must plan and account for. To ensure the lab's continuous and successful operation without requiring human intervention, the system should avoid any possible conflicts between these tasks while also maximizing the overall throughput by organizing the distribution of tasks in a parallel fashion.\\

\begin{figure}[htp]
    \centering
    \includegraphics[width=\linewidth]{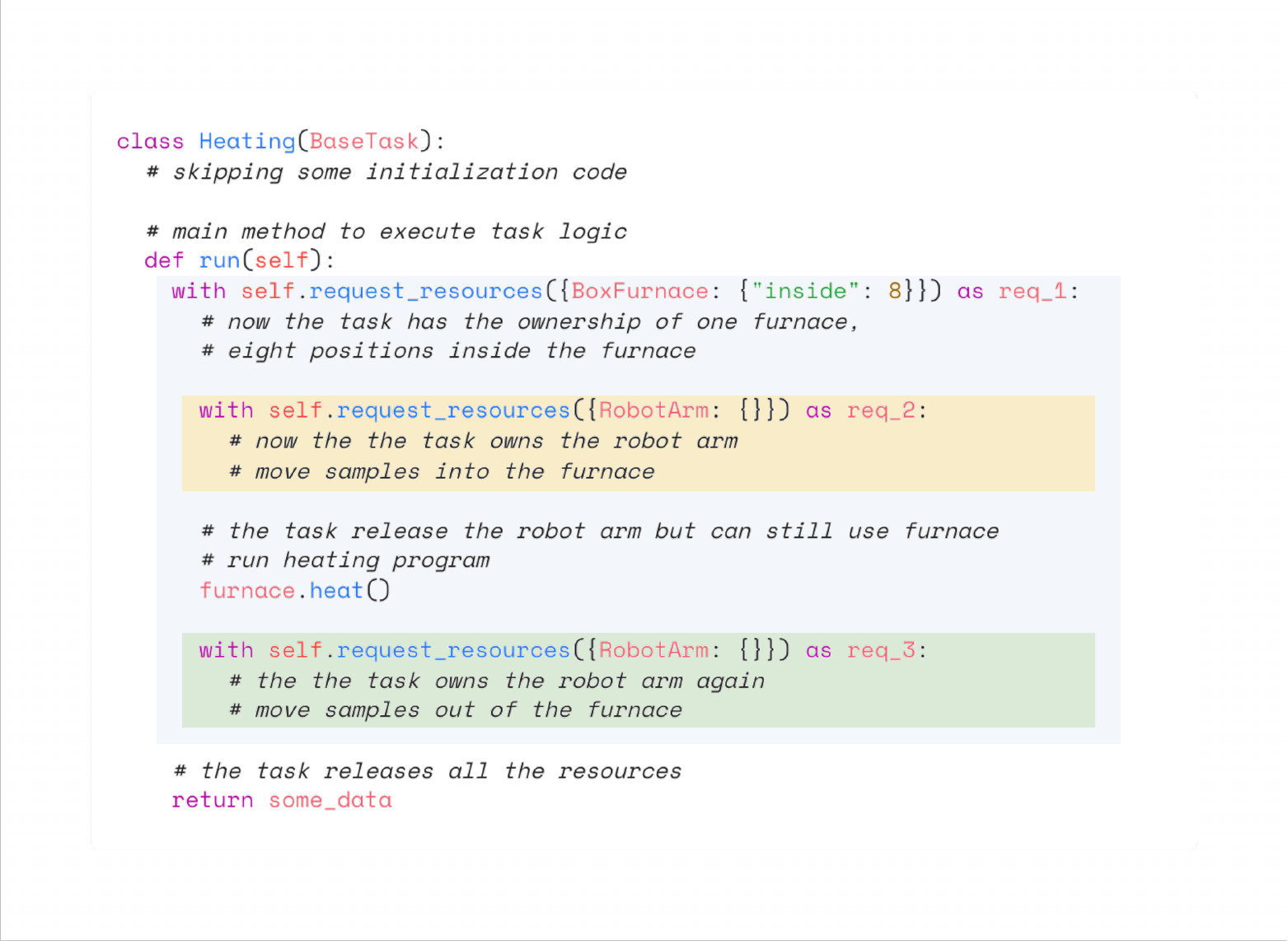}
    \caption{A code example of the resource request in the task definition. In this example, the heating task initiates three request resources with their contexts highlighted in different colors. The requested resources are assigned by the resource manager when entering Python's ``with'' context. It will be automatically released when exiting the context.}
    \label{fig:resource}
\end{figure}

In AlabOS, resource management is carried out using a cooperative multitasking schema \cite{jeffay1991non}. Before tasks perform any operations in the laboratory, it has to request the necessary resources from AlabOS system. All the resource requests are handled by a module named resource manager, which will parse each request and check if there are any idle devices and sample positions that can fulfill the request. Once the request can be fulfilled, the resource manager will mark the assigned devices and sample positions as occupied and let the task know which devices and sample positions it can use. The task can then send commands to the assigned devices and move samples into the assigned sample positions. Once those operations are finished, the associated resources are released to be assigned to other tasks. To avoid the occurrence of ``dead resources'' where tasks are completed without releasing their resources properly, the system introduces a with-statement context for all requests, as shown in Fig.~\ref{fig:resource}. Once resources are assigned to a task, they are recorded in a resource request context. All operations involving these resources need to be performed within this context to avoid a permission error. When the task is completed and exits the request context, it will automatically release all the requested resources, regardless of the task's success or failure.

A resource request can contain multiple devices and sample positions. In an autonomous laboratory, it is sometimes common to have multiple devices operating with the same functions (e.g., multiple furnaces for heating) to increase the lab's throughput. A task can operate on equivalent devices to obtain the desired results. In the task definition, a request can either specify the name of an exact device to use (e.g., Furnace A) or a type of device (e.g., muffle furnace). If a device type is specified, the resource manager will find any available device(s) under that type. The sample position request is then attached to the specific device. For example, one task may want to reserve the sample positions inside one furnace chamber. In this case, the sample positions request will be handled once the device request can be satisfied. When the resource manager finds an available combination of devices that can satisfy the request, it will advance to solve the available sample positions based on the proposed devices. Once both the devices and sample positions requests can be met, a request will be fulfilled and assigned to the corresponding tasks. With such a non-preemptive resource allocation strategy, the resource request serves in a ``first-come-first-serve'' strategy with the task priority taken into consideration. This does not need additional information like how long the task will occupy these resources. However, it also assumes that all the tasks do not occupy resources for an unreasonably long time which would lead to sample pileups in the lab. It is generally recommended for the laboratory developer to have a good knowledge of the bottlenecks and throughput of the laboratory and set up the right amount of tools to avoid this. On the other hand, it is always a good idea to break the resource request into many small pieces to avoid booking some resources for too long a time. Before the task definition is deployed in the physical laboratory, the lab developer should fully test the workflow definition in the simulation mode described in Section \ref{device_control} to ensure the expected behavior of a task definition. More sophisticated scheduling of equipment reservation techniques \cite{zhou2024multi} could be integrated within AlabOS to further optimize the throughput.\\

When initiating a resource request, a priority can also be attached to determine the urgency of the request. The resource request's priority is kept the same as the priority of the task, which is specified by the user during submission as a task parameter (default to ``NORMAL'' priority). The resource requests' priority can also be overridden by supplying a customized value when creating a resource request inside the task. It is sometimes useful to force a higher priority to the post-processing operations in a task. For example, the sample must be unloaded from the X-ray diffractometer before the next sample can be loaded. In this case, it is necessary to prioritize the unload operation to avoid gridlock in the laboratory. The priority is encoded as an integer ranging from 1 to 100. The default priority of all the tasks is 20, which indicates normal priority. When the resource manager polls the pending requests, it will first rank them by their priority and then by their submission time. The request that has a higher priority and was submitted earlier will generally be handled first.\\

\begin{figure}[H]
    \centering
    \includegraphics[width=0.9\linewidth]{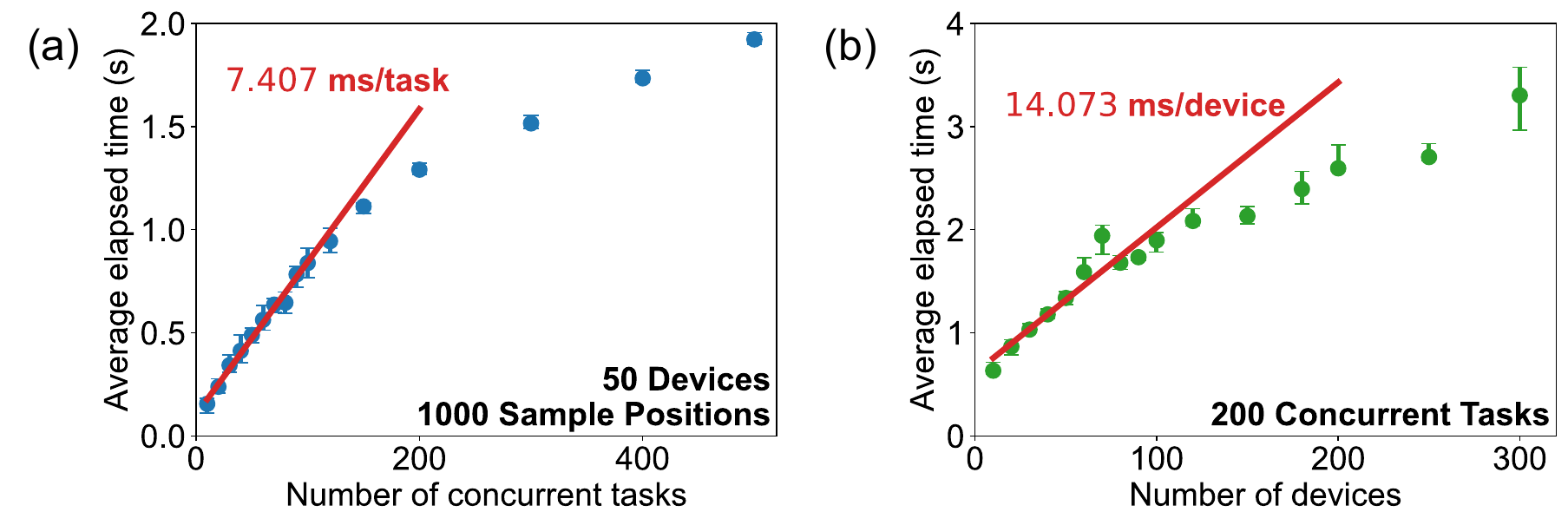}
    \caption{The execution time of processing resource requests under different numbers of tasks, devices, and sample positions. The time to process all the resource requests (a) with varied numbers of tasks in a virtual lab with 50 devices and 1000 sample positions (20 sample positions per device), (b) with fixed 200 tasks in a virtual lab with varied number of devices and sample positions (20 sample positions per device). The test is run on a Macbook Pro 2021 with M1 Pro CPU, 32 GB RAM. Each point is run 10 times repeatedly to obtain the error bar. The top and bottom of the error bar represent the 25\textsuperscript{th} and 75\textsuperscript{th} quantile of all the runtime data.}
    \label{fig:resource_bm}
\end{figure}

In an autonomous laboratory, many tasks can request resources at the same time. It is necessary to handle the large amount of possible resources on time so that the task can be executed faster. To understand the performance of the resource manager, a virtual lab with $M$ devices, each of which has 20 associated sample positions, is defined. At each step, there are $N$ tasks that request one device with one to twenty sample positions. Such configuration simulates the situation when the laboratory has a heavy workload. At each step, the resource manager will try to assign resources to each task. The CPU wall time is collected at different conditions. As shown in Figure.~\ref{fig:resource_bm}(a), when the number of devices is fixed, the processing time for all the tasks' requests scales linearly with the task number at small task numbers ($\leq$ 120). The average processing time for one task is 7.407 ms obtained from the slope of the fitted line. When the number of tasks continues to increase, the processing time deviates down from the fitted line. This may be attributed to a saturation of the resources. When there are way more tasks requesting resources compared to the available devices, most tasks' requests cannot be fulfilled in one step, leading to a processing time shorter than the time predicted by the linear relationship. The performance of the resource manager on different numbers of devices is shown in Figure.~\ref{fig:resource_bm}(b). Similar to the performance under varied task numbers, with the number of devices increasing, the processing time first follows a linear relationship with a slope of 14.073 ms/device and then becomes flat due to the saturation of the tasks. When the devices are much more than the tasks, most devices remain idle and do not cost any time for resource assignment.

Compared to the time required for completing a normal operation in the laboratory for solid-state synthesis, which usually takes several minutes to hours, the time for assigning resources is nearly negligible, thus ensuring a higher turnover rate in the laboratory. In some cases, multiple operations can be short and complete in a few seconds, while requiring different resources to complete. If that is the case, it is generally recommended to combine these short operations and request the resources all at once to minimize the waiting time.

\subsection{User interaction}
\begin{figure}[htp]
    \centering
    \includegraphics[width=\linewidth]{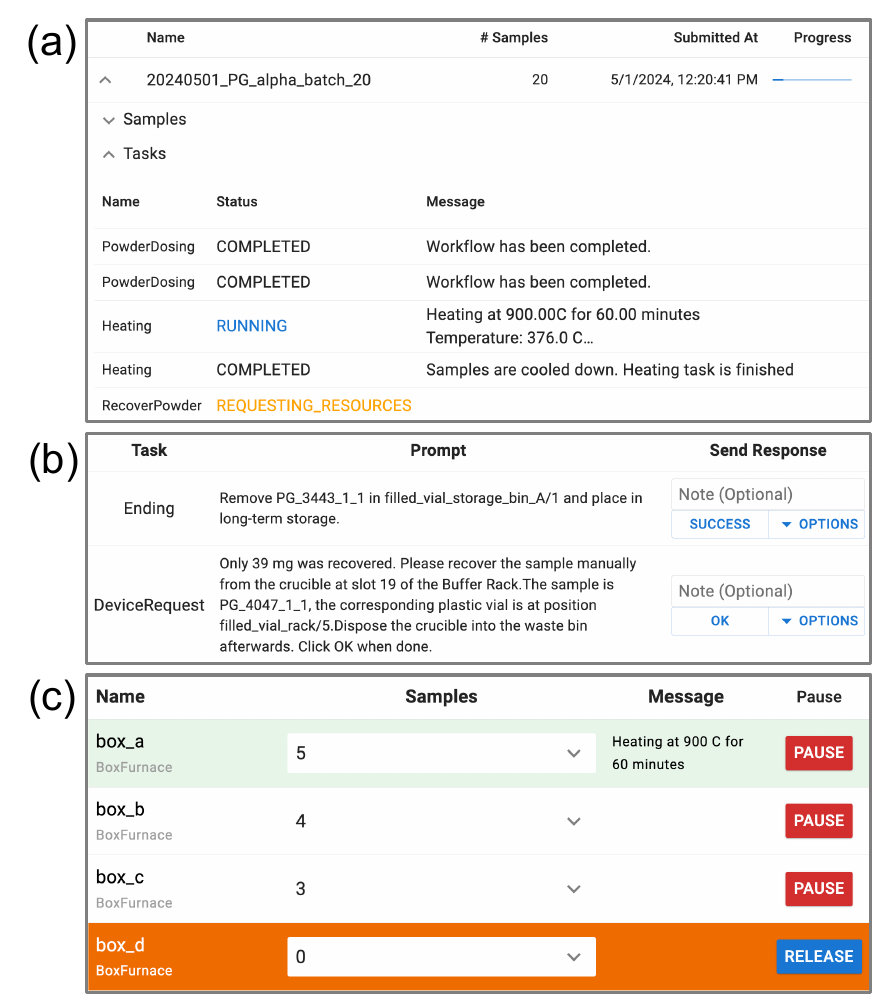}
    \caption{Images of three pages associated with different tabs in the dashboard. (a) The main tab shows the status of ongoing experiments. Each experiment has a progress bar with different colors, including blue (all tasks are running normally), green (all tasks are completed), and red (at least one task encountered an error). (b) The notification page shows messages from the system and its tasks. The operators can choose from the options provided and take action as instructed for maintenance and error recovery. (c) The device page shows the status of each device. When the operators wish to perform maintenance on a device, they can pause it from this page so that it will no longer be assigned to any tasks.}
    \label{fig:dashboard}
\end{figure}

To better facilitate daily maintenance and monitoring of experimental progress, AlabOS interacts with the operators through a web-based dashboard server (Fig.~\ref{fig:dashboard}). In the dashboard, the operators can check the real-time status of each sample, device, and task running or queued in the lab. It also provides a variety of interfaces for human-machine interaction. These are detailed below.

\subsubsection{Status monitoring}
The dashboard shows the real-time status of all experiments, tasks, devices, and samples running or queued in the lab (Fig.~\ref{fig:dashboard}a). An experiment will appear in the dashboard as soon as it is submitted to the system. The samples and tasks belonging to the experiment will also be shown in a scroll-down table. The sample position will be displayed next to each sample's name. Each task entry has a name, status, and message that provides human-readable information about what the task is currently doing. The message is designed by operators in the task definition with a \texttt{set\_message} method in the base task class. \\

In addition to monitoring the real-time status of the lab, the dashboard also provides a cancellation button for each ongoing experiment and task, which can force the unfinished experiments/tasks to stop running. When the task receives a cancellation signal but has not started, it will be canceled directly. If the task is running, an error will be raised in the task process, where it is automatically decided how to halt the process without causing a sudden interruption of the lab's robotic operation. 

\subsubsection{Notification}
Human interaction in an autonomous laboratory is essential to maintain a smooth operation. When something unexpected is detected in the laboratory (e.g., increased environment temperature, or the robot arm failing to pick up a sample), the autonomous lab should notify the operators in time to avoid further damage. Clear instructions should be given so that it requires minimal knowledge to act according to the notification.\\

In AlabOS, the notification can be initiated by either the system (manager instances) or the task processes. It can be used whenever a human action is needed. Each notification contains a prompt field and a list of possible options for the operators to select. The notification requires a receipt of acknowledgment from the operators with one option selected. Once that option is selected and the operator has completed the necessary tasks, the program will be carried out according to the selected actions. With clear notification messages designed when defining the tasks, the notification can be handled by the users without much knowledge about the autonomous laboratory. Thus, the maintenance jobs can be distributed among the lab developers and lab users.\\

Apart from the dashboard messages, notifications can also be configured to send over Slack bot or email. Further extensions can be made to other notification platforms, such as IFTTT \cite{mi2017empirical}, with minor programming efforts.

\subsubsection{Exception handling}
Because autonomous laboratories are generally used for research purposes, exceptions in hardware and software are inevitable. For example, the robot arm encountering an unexpected object can lead to a hardware exception, while the driver code failing to communicate with the robot arm would be considered a software exception. In AlabOS, errors are raised as standard Python exceptions in the task actor process. The exceptions are classified as recoverable or unrecoverable.
Recoverable exceptions refer to issues caught in the task process, usually in Python's ``try-except'' block. The raised error will not lead to the exiting of a program. Instead, there will be a handling routine that either retries or notifies human operators to check and recover manually. In contrast, an unrecoverable exception will lead to the failure of an entire task. These are not caught by the ``try-except'' block. AlabOS provides a default handling routine to notify human operators of all unrecoverable exceptions. They can then remove and replace any affected samples, reset the occupied devices, and resume the system to complete any other remaining samples.

\subsection{Device control} \label{device_control}
AlabOS system uses a centralized manager to monitor and control all devices in a laboratory. The task actors do not communicate directly with the physical equipment to execute commands and read data. Instead, they send all commands to the device manager through a remote procedure call (RPC) \cite{srinivasan1995rfc1831}. In each task, an identical RPC proxy class will be created as an in-place replacement for the task to send commands. In this way, the RPC communication will not change the way driver methods are called in the task definition, as opposed to directly calling the devices' methods.\\

Instead of creating the device driver objects in the task process, the indirect method call ensures the global singleton of each device. This helps to avoid the conflicts that can arise from multiple commands. Furthermore, the device manager checks the ownership of the requested devices and prevents them from being operated by tasks that do not reserve them. This minimizes the chance of conflicts and accidents in the laboratory.\\

To facilitate the rapid integration of new workflows in the autonomous laboratories, each device can be switched to simulation mode to allow tasks to run without connecting to the actual equipment. In the simulation mode, the operators can select to skip some methods (usually related to communication with the equipment) by applying a ``mock'' decorator to the method in the device definition. A mocked object with certain attributes will be specified as the return value in the ``mock'' decorator. In this way, the task can still proceed with the return values to test its functionalities, without requiring to talk with the hardware and their corresponding device drivers. 

\subsection{Data storage}
The system provides a logger object for tracking and storing data. Each piece of data is saved as a document in MongoDB, with some metadata that includes information such as the data source and its time of logging. The logged data is classified as device signal, sample amount, characterization result, system log, and ``other'' which includes all the data that do not fall into the previous four catalogs. The classification of logged data improves the organization of the database. Apart from the classification, each logged data also has a level to indicate its importance, ranging from 10 to 50. In this scheme, level 10 is intended for debugging, while level 50 is intended for fatal errors. In each task actor process, a logger object will be created at the beginning of the task. During runtime, it can be called conveniently to log the data. For example, it can be used to record the real-time temperature of the oven at a regular interval during the heating task, which can be read afterward from the database for debugging and post-analysis.\\

Despite the logger's ease of use, logged data can still become unstructured and scattered in the database. For example, each point of the real-time temperature of the furnace can be a single log document in the database. To make the data more accessible to researchers, we also provide a ``result'' field for each task. At the end of each task, the task can gather data generated during the run and return them as a Python dictionary. AlabOS stores all returned data in the result field. For example, one can return the temperature-time curve at the end of the heating task, which is especially useful when troubleshooting failed syntheses. If a large file needs to be stored in the database, AlabOS also provides a ``LargeResult'' object to handle the large file with MongoDB GridFS backend. The large file will be stored as chunks, with a reference ID generated and linked to the result collection.\\

When an experiment is completed, all data involving sample information, task results, and metadata can be copied to a backup database. This can be configured as a remote database on a cloud server. After being copied, no change will be made to these data. The operators can later query the data needed from this backup database. In this way, AlabOS ensures data safety and alleviates concerns about any potential file loss in the local database.

\subsection{Experiment Submission}
In AlabOS, all the experiments are submitted through the dashboard server, where an API is exposed to receive all the experiment submissions. The format of the experiment is first validated to ensure the correct processing afterward. The validated experiment is then dumped into an experiment collection, where the experiment manager instance keeps polling and builds samples and task graphs according to the experiment specifications. The raw input format is in JSON format, which includes the sample and task information. Although it is easy for the system to parse the JSON format, it is usually not straightforward for the users to compose the input file directly. An experiment builder class is introduced to help users to define experiments in a Python script. When an experiment builder is created, the name and metadata of the experiment should be provided. With an ``add\_sample'' method, the users can create and add samples to the experiments. Then the tasks are created and attached to the samples. The experiment builder will record the sequence of the tasks added to each sample and build the task graph with the dependency specified in the sequence. Finally, the experiment can be converted to a JSON format string and submitted to the submission API with a ``submit'' method. The users can also query the experiment status and results from the experiment information API.\\

The submission API and the experiment builder serve as a high-level abstraction to the autonomous laboratory. It resembles the queue management system in high-performance computing (HPC) clusters. When submitting jobs to HPC, the users do not need to pay attention to how and where their jobs are run. Similarly, when an experiment is submitted to AlabOS, the users can focus on designing experiments with proper parameters without knowing the details of how the experiment will be run. \\

The submission infrastructure of AlabOS opens up the possibility of conducting a close-loop experiment campaign inside an autonomous laboratory, where AI agents analyze the results and make plans for future experiments. With a structured input format that is sent to an API endpoint, there is no distinction between experiments submitted by humans and by AI. When designing a close-loop experiment campaign, the user can wrap the submission and result query script into an objective function with the inputs and outputs to be explored. The objective function can be directly used in various experiment planning software for self-driving laboratories like Chimera \cite{hase2018chimera} and Altas \cite{hickman_atlas_2023}.

\section{AlabOS in practice: orchestrating an autonomous laboratory for solid-state synthesis}
\begin{figure}[htp]
    \centering
    \includegraphics[width=\linewidth]{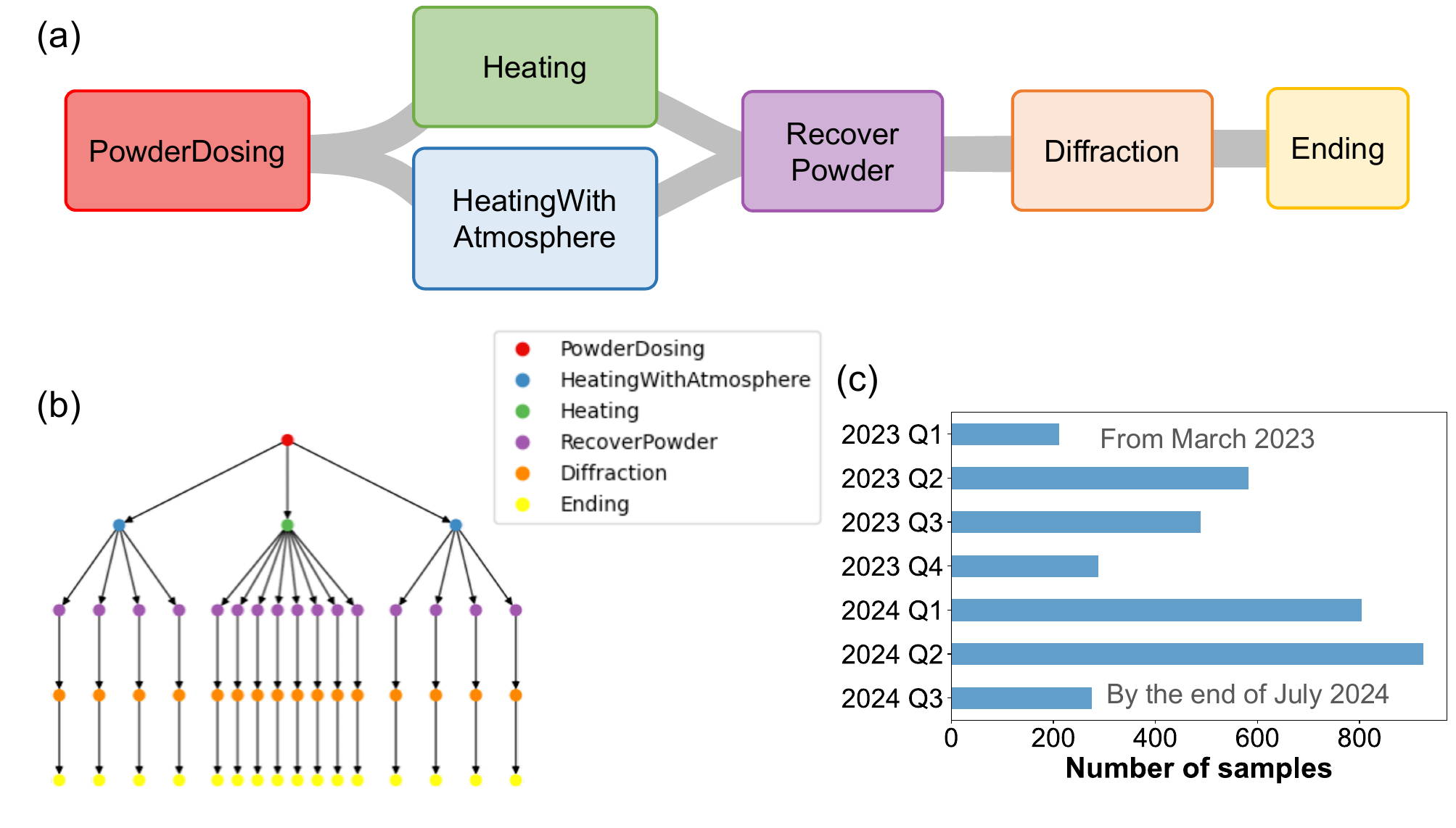}
    \caption{The typical tasks and workflows and sample throughput in the A-Lab. (a) In a typical workflow, one powder sample can be heated either in a box furnace (\texttt{Heating} task) or in a tube furnace where the atmosphere is controlled (\texttt{HeatingWithAtmosphere} task). (b) A batched task graph with 16 samples, including eight samples sent to the box furnace and eight to the tube furnace. (c) The number of distinct samples processed by A-Lab by quarter between March 2023 and July 2024.}
    \label{fig:a-lab}
\end{figure}

AlabOS is actively used to orchestrate the A-Lab \cite{Szymanski2023}, an autonomous synthesis platform for inorganic material synthesis. The A-Lab focuses specifically on preparing samples in powder form using solid-state synthesis and characterizes them using X-ray diffraction. A typical workflow of the A-Lab's operations is shown in Fig.~\ref{fig:a-lab}a. The workflow is divided into several tasks, which are the building blocks of each experiment in the A-Lab. A \texttt{PowderDosing} task weighs certain amounts of precursor powders and mixes them using ball-milling. A \texttt{Heating} task sends the samples into a box furnace, which is heated according to a pre-defined temperature profile. Similarly, the \texttt{HeatingWithAtmosphere} task uses tube furnaces to heat the samples under a controlled gas flow atmosphere. After the samples are heated, they are sent to the \texttt{RecoverPowder} task to be ground into fine powders. A small portion of the sample powder is extracted and sent to an X-ray diffractometer to measure the powder X-ray diffraction (XRD) spectrum in the \texttt{Diffraction} task. At the end of one experimental cycle, the \texttt{Ending} task notifies the operators to store each completed sample in a designated position, and this position is updated in the database accordingly.\\
 
To set up the A-Lab workflows using AlabOS, a project folder is first created. A \texttt{toml} format configuration file is made to store all the connection information to the database, the notification service, and the message queue service. The devices' driver codes are stored in a \texttt{devices} folder. Each type of device is defined in a Python object class inherited from \texttt{BaseDevice} to include some basic functions like connecting and status checking. Apart from these, we can define as many methods as needed to operate the device. We also define the associated sample position names in the device. They are read by AlabOS and recorded in the database when launching. To build an automated lab for solid-state synthesis and powder XRD characterization, there are in total 16 types of devices with 28 device instances and 289 device-related sample positions. The full list of defined devices, their communication protocols, the number of associated sample positions, and numbers of handled exceptions are listed in Table~.\ref{table:devices}. Most devices communicate with the control PC through Ethernet, with various protocols including HTTP, MODBUS, XML-RPC, etc. Each device (and control PC) is assigned to a unique IP within an intranet. The control PC can send commands to the devices by specifying the IP addresses of the devices. The box furnaces, as an exception, are connected directly to the control PC through its serial interfaces, with a COM port assigned to each furnace.\\

\newgeometry{margin=1in}
\begin{landscape}
\begin{table}[]
\centering
\resizebox{\linewidth}{!}{%
\begin{tabular}{|c|c|c|c|c|c|}
\hline
\textbf{Device Type} &
  \textbf{Physical Device} &
  \textbf{Communication Protocol} &
  \textbf{Number of Devices} &
  \textbf{Number of Associated Sample Positions Per Device}
\\ \hline
\texttt{BallDispenser}            & Arduino                                   & HTTP               & 1 & 1 \\ \hline
\texttt{BoxFurnace}               & Thermo Scientific F4805560                & MODBUS over Serial & 4 & 8 \\ \hline
\texttt{CapDispenser}             & Arduino                                   & HTTP               & 1 & 3 \\ \hline
\texttt{CappingGripper}           & Arduino                                   & HTTP               & 1 & 1 \\ \hline
\texttt{Diffractometer}           & Malvern Panalytical Aeris Mineral Edition & Socket             & 1 & 1 \\ \hline
\texttt{LabmanQuadrant}           & Labman Powder Dosing System               & HTTP               & 4 & 48 \\ \hline
\texttt{ManualFurnace}            & N/A                                       & N/A                & 4 & 8  \\ \hline
\texttt{RobotArmCharacterization} & Universal Robot 5e                        & Socket \& SSH      & 1 & 1  \\ \hline
\texttt{RobotArmFurnaces}         & Universal Robot 5e                        & Socket \& SSH      & 1 & 1  \\ \hline
\texttt{Scale}                    & OHAUS Scout                               & HTTP               & 1 & 1  \\ \hline
\texttt{Shaker}                   & Arduino                                   & HTTP               & 1 & 1  \\ \hline
\texttt{TransferRack}             & N/A                                       & N/A                & 1 & 6  \\ \hline
\texttt{TubeFurnace}              & MTI OTF1200X5ASD                          & XML-RPC            & 4 & 4  \\ \hline
\texttt{VialDispenserRack}        & N/A                                       & N/A                & 1 & 0  \\ \hline
\texttt{VialLabeler}              & Reiner jetStamp 1025                      & Serial             & 1 & 1  \\ \hline
\texttt{XRDDispenserRack}         & N/A                                       & N/A                & 1 & 0  \\ \hline
\end{tabular}%
}
\caption{All the devices implemented in A-Lab within AlabOS framework. Each device type represents a Python class definition using AlabOS APIs. The physical devices used in A-Lab are listed. The communication protocol indicates how the device object communicates with the physical device in A-Lab. If the protocol is N/A, it indicates the device is not linked to any physical devices and only exists in AlabOS system to manage some status of the lab. For example, the \texttt{XRDDispenserRack} is used to monitor the number of clean XRD sample holders and will notify users when there are no more clean XRD holders. The number of devices and the associated sample positions for each device are also listed. }
\label{table:devices}
\end{table}
\end{landscape}
\restoregeometry

After the devices are set, another folder named \texttt{tasks} is created to store all the task procedures. Each task inherits from \texttt{BaseTask} in AlabOS, which implements methods to report progress, reserve resources, and get samples' information. In each task, the device and sample position resources are reserved before any actions are taken in A-Lab. A device handle is returned after the resource is assigned in AlabOS. The task can call any method in the device class to perform operations. Finally, the system can be launched via a terminal's command line interface (CLI). If any changes are made to the task definition, the system will need to be restarted to take the changes into effect.\\

To optimize throughput, several tasks in the A-Lab can handle multiple samples, with the capacity of each task and the number of successfully completed tasks are shown in Table~.\ref{table-capacity}. For example, samples are loaded into a ceramic rack with eight slots in the heating task. A robot arm then inserts this entire ceramic rack into a box furnace. Such batch processing approach gives the heating task a maximum capacity of eight. When submitting the experiment, one node in the task graph can take multiple samples, which indicates that they are processed in the same batch. For example, an experiment with sixteen samples has the task graph shown in Fig.~\ref{fig:a-lab}b. All of the sixteen samples are processed in one powder dosing task. The samples are then divided into three heating batches with four, four, and eight samples, respectively. These are heated in different furnaces according to the specified heating profiles. Samples with different heating profiles cannot share the same furnace. \\

\begin{table}[]
\resizebox{\textwidth}{!}{%
\begin{tabular}{|c|c|c|c|c|c|}
\hline
  \textbf{Task name} &
  \textbf{Capacity} &
  \textbf{Number of tasks} &
  \textbf{Error handling routines in definition} &
  \textbf{\% with exceptions} &
  \textbf{\% with unrecoverable exceptions} \\ \hline
\texttt{PowderDosing}          & 16 & 333  & 1  & 1.20  & 1.20 \\ \hline
\texttt{Heating}               & 8  & 403  & 2  & 4.96  & 4.22 \\ \hline
\texttt{ManualHeating}         & 8  & 75   & 0  & 0     & 0    \\ \hline
\texttt{HeatingWithAtmosphere} & 4  & 27   & 0  & 22.22 & 7.41 \\ \hline
\texttt{RecoverPowder}         & 1  & 2818 & 15 & 4.19  & 0.99 \\ \hline
\texttt{Diffraction}           & 1  & 2658 & 11 & 4.74  & 1.20 \\ \hline
\texttt{Ending}                & 1  & 2574 & 3  & 0.12  & 0.04 \\ \hline
\end{tabular}%
}
\caption{The maximum capacity of each task and the number of tasks in A-Lab by the time of writing. The number of handled errors in each task, indicated by the try/except blocks with user inputs requested, is shown in the table. Among all the tasks, the ratios of tasks with exceptions and with unrecoverable exceptions are also shown in the last two columns.}
\label{table-capacity}
\end{table}

While a common workflow of tasks is shown in Fig.~\ref{fig:a-lab}a, many other workflows are possible in the A-Lab. Several of these are shown in Table~\ref{table:others}. For example, when the operators wish to heat samples manually while still using the rest of the automated processes in the A-Lab, a \texttt{ManualHeating} task can inform the operators (via the notification system) of the position of each sample after the powder dosing task is completed. It will also notify the operators where to place the heated samples so that any downstream tasks can process them. As another example, if the operators wish to obtain the XRD pattern of precursors used in the A-Lab, they can skip the heating tasks and move the sample (precursor powders) directly to the powder recovery and XRD tasks. Also, if the operators would like to process a set of samples that are heated outside A-Lab, they can use a \texttt{Starting} task to create the record of the sample in the database and specify the position of the sample so that tasks can find the sample.
\begin{table}[H]
\centering
\begin{tabular}{|c|c|c|}
\hline
  & \textbf{Workflow}                                                                                                       & \textbf{Note}                                                                                                       \\ \hline
1 & \begin{tabular}[c]{@{}c@{}}\texttt{PowderDosing} - \texttt{ManualHeating} - \\ \texttt{RecoverPowder} - \texttt{Diffraction} - \texttt{Ending}\end{tabular} & \begin{tabular}[c]{@{}c@{}}Heat the samples in the external \\ furnaces for higher throughput\end{tabular} \\ \hline
2 & \begin{tabular}[c]{@{}c@{}}\texttt{PowderDosing} - \texttt{RecoverPowder} - \\ \texttt{Diffraction} - \texttt{Ending}\end{tabular}                 & \begin{tabular}[c]{@{}c@{}}Skip the heating to characterize\\ the precursors\end{tabular}                  \\ \hline
3 & \begin{tabular}[c]{@{}c@{}}\texttt{Starting} - \texttt{RecoverPowder} - \\ \texttt{Diffraction} - \texttt{Ending}\end{tabular}                     & \begin{tabular}[c]{@{}c@{}}Process samples that are made\\ outside A-Lab\end{tabular}                      \\ \hline
\end{tabular}%
\caption{Other frequently used workflows in A-Lab. The workflow column shows the sequence of the tasks. The note column describes the use cases of each workflow.}
\label{table:others}
\end{table}

By the time of writing, AlabOS has been driving A-Lab to synthesize and characterize over 3,500 distinct samples over a period of approximately one and half years, as shown in Fig~.\ref{fig:a-lab}c. The maximum number of samples submitted in one single day was 149 samples on Feb 9, 2024. The system is designed to be able to handle a large amount of sample submissions at a time. By dividing the experiments composed of many samples into modular tasks, AlabOS is able to schedule and manage experiments at a finer granularity, thus ensuring the high efficiency of completing the complicated workflows.

To maintain smooth operation, A-Lab task definitions incorporate handling routines for common exceptions. The number of exceptions in each A-Lab task is listed in Table.~\ref{table-capacity}. Depending on the complexity of the task, different exception-handling codes are inserted into task definitions. Among them, \texttt{RecoverPowder} and \texttt{Diffraction} require the most exception handling as many steps are involved in the operation. The number of exception-handling routines also aligns with the ratio of recovered exceptions in all the exceptions raised, as shown in Table.~\ref{table-capacity}. \texttt{RecoverPowder} and \texttt{Diffraction} demonstrate a higher exception recovery rate of 76\% and 75\%, respectively. On the contrary, \texttt{Heating} and \texttt{PowderDosing} have a lower recovery rate of 15\% and 0\% due to limited exception handling. The lack of routines to address exceptions in these tasks can be attributed to the fact that some are due to hardware that is worn out, or to newly emerged software communication exceptions, which were not detected in the soak test. Over time, these tasks will suffer less from unrecoverable errors in the future with more exception-handling routines included. 

\section{Conclusion and Outlook}
We have outlined the development and application of AlabOS as an orchestration software for managing workflows in autonomous laboratories. This package is designed to meet the fast-changing requirements of a typical materials research program. With the graph-based experiment input format, independent task actor design, and the resource occupation mechanism, human operators can define and submit workflows in AlabOS without concerning themselves with possible conflicts between concurrent tasks. The dashboard and notification system provide a general solution for human-machine interaction in the laboratory, which becomes especially useful when performing maintenance and error recovery jobs.
With the public availability of AlabOS, we hope that researchers will no longer need to write complex workflow management codes but instead focus their time on developing high-level logic followed by autonomous laboratories for accelerated materials discovery and optimization. Only in this way, the autonomous laboratory can be more accessible to general researchers and become a more powerful tool for accelerating material discovery.

\section{Code availability}
The code for the AlabOS is available at \url{https://github.com/CederGroupHub/alabos}. The version of the AlabOS code employed for this study is v1.0.1. The online document is hosted at \url{https://cedergrouphub.github.io/alabos/}. The implementation of device communication driver is available at \url{https://github.com/CederGroup/alab_control}. The A-Lab-specific device and task definitions are available at \url{https://github.com/CederGroupHub/alabos/tree/main/examples/alab_example}. 

\section{Author Contributions}
Y.F. and B.R.: conceptualization, software, writing - original draft, writing - review and editing. R.K.: conceptualization, software, writing - review and editing. O.D., H.P.S., and M.J.M: software, writing - review and editing. Z.W., N.J.S., L.N.W., and D.M: investigation, writing - review and editing. Y.Z.: methodology, supervision, writing - review and editing. A.J.: methodology, supervision, resources, writing - review and editing. G.C.: resources, supervision, methodology, project administration, writing - review and editing.

\section{Conflicts of interest}
There are no conflicts to declare.

\section{Acknowledgements}
This work was primarily financed by the U.S. Department of Energy, Office of Science, Office of Basic Energy Sciences, Materials Sciences and Engineering Division under contract no. DE-AC02-05-CH11231 (D2S2 programme, KCD2S2), the Laboratory Directed Research and Development Program of Lawrence Berkeley National Laboratory, and Materials Project. Work done at UC Berkeley was supported by Umicore Specialty Oxides and Chemicals.

\bibliography{rsc}

\providecommand*{\mcitethebibliography}{\thebibliography}
\csname @ifundefined\endcsname{endmcitethebibliography}
{\let\endmcitethebibliography\endthebibliography}{}
\begin{mcitethebibliography}{60}
\providecommand*{\natexlab}[1]{#1}
\providecommand*{\mciteSetBstSublistMode}[1]{}
\providecommand*{\mciteSetBstMaxWidthForm}[2]{}
\providecommand*{\mciteBstWouldAddEndPuncttrue}
  {\def\EndOfBibitem{\unskip.}}
\providecommand*{\mciteBstWouldAddEndPunctfalse}
  {\let\EndOfBibitem\relax}
\providecommand*{\mciteSetBstMidEndSepPunct}[3]{}
\providecommand*{\mciteSetBstSublistLabelBeginEnd}[3]{}
\providecommand*{\EndOfBibitem}{}
\mciteSetBstSublistMode{f}
\mciteSetBstMaxWidthForm{subitem}
{(\emph{\alph{mcitesubitemcount}})}
\mciteSetBstSublistLabelBeginEnd{\mcitemaxwidthsubitemform\space}
{\relax}{\relax}

\bibitem[Jain \emph{et~al.}(2013)Jain, Ong, Hautier, Chen, Richards, Dacek, Cholia, Gunter, Skinner, Ceder, and Persson]{jain2013commentary}
A.~Jain, S.~P. Ong, G.~Hautier, W.~Chen, W.~D. Richards, S.~Dacek, S.~Cholia, D.~Gunter, D.~Skinner, G.~Ceder and K.~A. Persson, \emph{APL Materials}, 2013, \textbf{1}, 011002\relax
\mciteBstWouldAddEndPuncttrue
\mciteSetBstMidEndSepPunct{\mcitedefaultmidpunct}
{\mcitedefaultendpunct}{\mcitedefaultseppunct}\relax
\EndOfBibitem
\bibitem[Schmidt \emph{et~al.}(2019)Schmidt, Marques, Botti, and Marques]{schmidt2019recent}
J.~Schmidt, M.~R.~G. Marques, S.~Botti and M.~A.~L. Marques, \emph{npj Computational Materials}, 2019, \textbf{5}, 83\relax
\mciteBstWouldAddEndPuncttrue
\mciteSetBstMidEndSepPunct{\mcitedefaultmidpunct}
{\mcitedefaultendpunct}{\mcitedefaultseppunct}\relax
\EndOfBibitem
\bibitem[Choudhary \emph{et~al.}(2022)Choudhary, DeCost, Chen, Jain, Tavazza, Cohn, Park, Choudhary, Agrawal, Billinge, Holm, Ong, and Wolverton]{choudhary2022recent}
K.~Choudhary, B.~DeCost, C.~Chen, A.~Jain, F.~Tavazza, R.~Cohn, C.~W. Park, A.~Choudhary, A.~Agrawal, S.~J.~L. Billinge, E.~Holm, S.~P. Ong and C.~Wolverton, \emph{npj Computational Materials}, 2022, \textbf{8}, 59\relax
\mciteBstWouldAddEndPuncttrue
\mciteSetBstMidEndSepPunct{\mcitedefaultmidpunct}
{\mcitedefaultendpunct}{\mcitedefaultseppunct}\relax
\EndOfBibitem
\bibitem[Butler \emph{et~al.}(2018)Butler, Davies, Cartwright, Isayev, and Walsh]{butler2018machine}
K.~T. Butler, D.~W. Davies, H.~Cartwright, O.~Isayev and A.~Walsh, \emph{Nature}, 2018, \textbf{559}, 547--555\relax
\mciteBstWouldAddEndPuncttrue
\mciteSetBstMidEndSepPunct{\mcitedefaultmidpunct}
{\mcitedefaultendpunct}{\mcitedefaultseppunct}\relax
\EndOfBibitem
\bibitem[Schleder \emph{et~al.}(2019)Schleder, Padilha, Acosta, Costa, and Fazzio]{schleder2019dft}
G.~R. Schleder, A.~C. Padilha, C.~M. Acosta, M.~Costa and A.~Fazzio, \emph{Journal of Physics: Materials}, 2019, \textbf{2}, 032001\relax
\mciteBstWouldAddEndPuncttrue
\mciteSetBstMidEndSepPunct{\mcitedefaultmidpunct}
{\mcitedefaultendpunct}{\mcitedefaultseppunct}\relax
\EndOfBibitem
\bibitem[Morgan and Jacobs(2020)]{morgan2020opportunities}
D.~Morgan and R.~Jacobs, \emph{Annual Review of Materials Research}, 2020, \textbf{50}, 71--103\relax
\mciteBstWouldAddEndPuncttrue
\mciteSetBstMidEndSepPunct{\mcitedefaultmidpunct}
{\mcitedefaultendpunct}{\mcitedefaultseppunct}\relax
\EndOfBibitem
\bibitem[Kirklin \emph{et~al.}(2015)Kirklin, Saal, Meredig, Thompson, Doak, Aykol, R{\"u}hl, and Wolverton]{kirklin2015open}
S.~Kirklin, J.~E. Saal, B.~Meredig, A.~Thompson, J.~W. Doak, M.~Aykol, S.~R{\"u}hl and C.~Wolverton, \emph{npj Computational Materials}, 2015, \textbf{1}, 1--15\relax
\mciteBstWouldAddEndPuncttrue
\mciteSetBstMidEndSepPunct{\mcitedefaultmidpunct}
{\mcitedefaultendpunct}{\mcitedefaultseppunct}\relax
\EndOfBibitem
\bibitem[Sumpter \emph{et~al.}(2015)Sumpter, Vasudevan, Potok, and Kalinin]{SumpterCNPGM}
B.~G. Sumpter, R.~K. Vasudevan, T.~Potok and S.~V. Kalinin, \emph{npj Computational Materials}, 2015, \textbf{1}, 15008\relax
\mciteBstWouldAddEndPuncttrue
\mciteSetBstMidEndSepPunct{\mcitedefaultmidpunct}
{\mcitedefaultendpunct}{\mcitedefaultseppunct}\relax
\EndOfBibitem
\bibitem[Chamorro and McQueen(2018)]{ChamorroACR}
J.~R. Chamorro and T.~M. McQueen, \emph{Accounts of Chemical Research}, 2018, \textbf{51}, 2918--2925\relax
\mciteBstWouldAddEndPuncttrue
\mciteSetBstMidEndSepPunct{\mcitedefaultmidpunct}
{\mcitedefaultendpunct}{\mcitedefaultseppunct}\relax
\EndOfBibitem
\bibitem[Wang \emph{et~al.}(2024)Wang, Sun, Cruse, Zeng, Fei, Liu, Shangguan, Byeon, Jun, He, Sun, and Ceder]{WangNatSynth}
Z.~Wang, Y.~Sun, K.~Cruse, Y.~Zeng, Y.~Fei, Z.~Liu, J.~Shangguan, Y.-W. Byeon, K.~Jun, T.~He, W.~Sun and G.~Ceder, \emph{Nature Synthesis}, 2024\relax
\mciteBstWouldAddEndPuncttrue
\mciteSetBstMidEndSepPunct{\mcitedefaultmidpunct}
{\mcitedefaultendpunct}{\mcitedefaultseppunct}\relax
\EndOfBibitem
\bibitem[Steiner \emph{et~al.}(2019)Steiner, Wolf, Glatzel, Andreou, Granda, Keenan, Hinkley, Aragon-Camarasa, Kitson, Angelone, and Cronin]{steiner2019organic}
S.~Steiner, J.~Wolf, S.~Glatzel, A.~Andreou, J.~M. Granda, G.~Keenan, T.~Hinkley, G.~Aragon-Camarasa, P.~J. Kitson, D.~Angelone and L.~Cronin, \emph{Science}, 2019, \textbf{363}, eaav2211\relax
\mciteBstWouldAddEndPuncttrue
\mciteSetBstMidEndSepPunct{\mcitedefaultmidpunct}
{\mcitedefaultendpunct}{\mcitedefaultseppunct}\relax
\EndOfBibitem
\bibitem[Hartrampf \emph{et~al.}(2020)Hartrampf, Saebi, Poskus, Gates, Callahan, Cowfer, Hanna, Antilla, Schissel, Quartararo,\emph{et~al.}]{hartrampf2020synthesis}
N.~Hartrampf, A.~Saebi, M.~Poskus, Z.~P. Gates, A.~J. Callahan, A.~E. Cowfer, S.~Hanna, S.~Antilla, C.~K. Schissel, A.~J. Quartararo \emph{et~al.}, \emph{Faculty Opinions – Post-Publication Peer Review of the Biomedical Literature}, 2020, \textbf{368}, 980--987\relax
\mciteBstWouldAddEndPuncttrue
\mciteSetBstMidEndSepPunct{\mcitedefaultmidpunct}
{\mcitedefaultendpunct}{\mcitedefaultseppunct}\relax
\EndOfBibitem
\bibitem[Manzano \emph{et~al.}(2022)Manzano, Hou, Frei, Wang, Kitson, and Cronin]{manzano2022autonomous}
J.~S. Manzano, S.~S. Hou, Wenduan a†d~Zalesskiy, P.~Frei, H.~Wang, P.~J. Kitson and L.~Cronin, \emph{Nature Chemistry}, 2022, \textbf{14}, 1311--1318\relax
\mciteBstWouldAddEndPuncttrue
\mciteSetBstMidEndSepPunct{\mcitedefaultmidpunct}
{\mcitedefaultendpunct}{\mcitedefaultseppunct}\relax
\EndOfBibitem
\bibitem[Bennett \emph{et~al.}(2024)Bennett, Orouji, Khan, Sadeghi, Rodgers, and Abolhasani]{bennett2024autonomous}
J.~Bennett, N.~Orouji, M.~Khan, S.~Sadeghi, J.~Rodgers and M.~Abolhasani, \emph{Nature Chemical Engineering}, 2024,  1--11\relax
\mciteBstWouldAddEndPuncttrue
\mciteSetBstMidEndSepPunct{\mcitedefaultmidpunct}
{\mcitedefaultendpunct}{\mcitedefaultseppunct}\relax
\EndOfBibitem
\bibitem[MacLeod \emph{et~al.}(2021)MacLeod, Parlane, Morrissey, H{\"a}se, Roch, Dettelbach, Moreira, Yunker, Rooney, Deeth,\emph{et~al.}]{macleod2020self}
B.~P. MacLeod, F.~G. Parlane, T.~D. Morrissey, F.~H{\"a}se, L.~M. Roch, K.~E. Dettelbach, R.~Moreira, L.~P. Yunker, M.~B. Rooney, J.~R. Deeth \emph{et~al.}, \emph{Chem}, 2021, \textbf{7}, 2541--2545\relax
\mciteBstWouldAddEndPuncttrue
\mciteSetBstMidEndSepPunct{\mcitedefaultmidpunct}
{\mcitedefaultendpunct}{\mcitedefaultseppunct}\relax
\EndOfBibitem
\bibitem[Kusne \emph{et~al.}(2020)Kusne, Yu, Wu, Zhang, Hattrick-Simpers, DeCost, Sarker, Oses, Toher, Curtarolo, Davydov, Agarwal, Bendersky, Li, Mehta, and Takeuchi]{kusne2020fly}
A.~G. Kusne, H.~Yu, C.~Wu, H.~Zhang, J.~Hattrick-Simpers, B.~DeCost, S.~Sarker, C.~Oses, C.~Toher, S.~Curtarolo, A.~V. Davydov, R.~Agarwal, L.~A. Bendersky, M.~Li, A.~Mehta and I.~Takeuchi, \emph{Nature Communications}, 2020, \textbf{11}, 5966\relax
\mciteBstWouldAddEndPuncttrue
\mciteSetBstMidEndSepPunct{\mcitedefaultmidpunct}
{\mcitedefaultendpunct}{\mcitedefaultseppunct}\relax
\EndOfBibitem
\bibitem[Burger \emph{et~al.}(2020)Burger, Maffettone, Gusev, Aitchison, Bai, Wang, Li, Alston, Li, Clowes,\emph{et~al.}]{burger2020mobile}
B.~Burger, P.~M. Maffettone, V.~V. Gusev, C.~M. Aitchison, Y.~Bai, X.~Wang, X.~Li, B.~M. Alston, B.~Li, R.~Clowes \emph{et~al.}, \emph{Nature}, 2020, \textbf{583}, 237--241\relax
\mciteBstWouldAddEndPuncttrue
\mciteSetBstMidEndSepPunct{\mcitedefaultmidpunct}
{\mcitedefaultendpunct}{\mcitedefaultseppunct}\relax
\EndOfBibitem
\bibitem[Szymanski \emph{et~al.}(2023)Szymanski, Rendy, Fei, Kumar, He, Milsted, McDermott, Gallant, Cubuk, Merchant, Kim, Jain, Bartel, Persson, Zeng, and Ceder]{Szymanski2023}
N.~J. Szymanski, B.~Rendy, Y.~Fei, R.~E. Kumar, T.~He, D.~Milsted, M.~J. McDermott, M.~Gallant, E.~D. Cubuk, A.~Merchant, H.~Kim, A.~Jain, C.~J. Bartel, K.~Persson, Y.~Zeng and G.~Ceder, \emph{Nature}, 2023\relax
\mciteBstWouldAddEndPuncttrue
\mciteSetBstMidEndSepPunct{\mcitedefaultmidpunct}
{\mcitedefaultendpunct}{\mcitedefaultseppunct}\relax
\EndOfBibitem
\bibitem[Chen \emph{et~al.}(2024)Chen, Cross, Miara, Cho, Wang, and Sun]{chen2024navigating}
J.~Chen, S.~R. Cross, L.~J. Miara, J.-J. Cho, Y.~Wang and W.~Sun, \emph{Nature Synthesis}, 2024,  1--9\relax
\mciteBstWouldAddEndPuncttrue
\mciteSetBstMidEndSepPunct{\mcitedefaultmidpunct}
{\mcitedefaultendpunct}{\mcitedefaultseppunct}\relax
\EndOfBibitem
\bibitem[Correa-Baena \emph{et~al.}(2018)Correa-Baena, Hippalgaonkar, van Duren, Jaffer, Chandrasekhar, Stevanovic, Wadia, Guha, and Buonassisi]{correa2018accelerating}
J.-P. Correa-Baena, K.~Hippalgaonkar, J.~van Duren, S.~Jaffer, V.~R. Chandrasekhar, V.~Stevanovic, C.~Wadia, S.~Guha and T.~Buonassisi, \emph{Joule}, 2018, \textbf{2}, 1410--1420\relax
\mciteBstWouldAddEndPuncttrue
\mciteSetBstMidEndSepPunct{\mcitedefaultmidpunct}
{\mcitedefaultendpunct}{\mcitedefaultseppunct}\relax
\EndOfBibitem
\bibitem[Saal \emph{et~al.}(2013)Saal, Kirklin, Aykol, Meredig, and Wolverton]{saal2013materials}
J.~E. Saal, S.~Kirklin, M.~Aykol, B.~Meredig and C.~Wolverton, \emph{Jom}, 2013, \textbf{65}, 1501--1509\relax
\mciteBstWouldAddEndPuncttrue
\mciteSetBstMidEndSepPunct{\mcitedefaultmidpunct}
{\mcitedefaultendpunct}{\mcitedefaultseppunct}\relax
\EndOfBibitem
\bibitem[Liu \emph{et~al.}(2021)Liu, Zhao, Liu, Xi, Li, Xiang, and Zhou]{liu2021application}
B.~Liu, J.~Zhao, Y.~Liu, J.~Xi, Q.~Li, H.~Xiang and Y.~Zhou, \emph{Journal of Materials Science \& Technology}, 2021, \textbf{88}, 143--157\relax
\mciteBstWouldAddEndPuncttrue
\mciteSetBstMidEndSepPunct{\mcitedefaultmidpunct}
{\mcitedefaultendpunct}{\mcitedefaultseppunct}\relax
\EndOfBibitem
\bibitem[Szymanski \emph{et~al.}(2021)Szymanski, Bartel, Zeng, Tu, and Ceder]{szymanski2021probabilistic}
N.~J. Szymanski, C.~J. Bartel, Y.~Zeng, Q.~Tu and G.~Ceder, \emph{Chemistry of Materials}, 2021, \textbf{33}, 4204--4215\relax
\mciteBstWouldAddEndPuncttrue
\mciteSetBstMidEndSepPunct{\mcitedefaultmidpunct}
{\mcitedefaultendpunct}{\mcitedefaultseppunct}\relax
\EndOfBibitem
\bibitem[Stanev \emph{et~al.}(2018)Stanev, Vesselinov, Kusne, Antoszewski, Takeuchi, and Alexandrov]{stanev2018unsupervised}
V.~Stanev, V.~V. Vesselinov, A.~G. Kusne, G.~Antoszewski, I.~Takeuchi and B.~S. Alexandrov, \emph{npj Computational Materials}, 2018, \textbf{4}, 43\relax
\mciteBstWouldAddEndPuncttrue
\mciteSetBstMidEndSepPunct{\mcitedefaultmidpunct}
{\mcitedefaultendpunct}{\mcitedefaultseppunct}\relax
\EndOfBibitem
\bibitem[Chen \emph{et~al.}(2021)Chen, Bai, Ament, Zhao, Guevarra, Zhou, Selman, van Dover, Gregoire, and Gomes]{chen2021automating}
D.~Chen, Y.~Bai, S.~Ament, W.~Zhao, D.~Guevarra, L.~Zhou, B.~Selman, R.~B. van Dover, J.~M. Gregoire and C.~P. Gomes, \emph{Nature Machine Intelligence}, 2021, \textbf{3}, 812--822\relax
\mciteBstWouldAddEndPuncttrue
\mciteSetBstMidEndSepPunct{\mcitedefaultmidpunct}
{\mcitedefaultendpunct}{\mcitedefaultseppunct}\relax
\EndOfBibitem
\bibitem[Liu \emph{et~al.}(2017)Liu, Osadchy, Ashton, Foster, Solomon, and Gibson]{liu2017deep}
J.~Liu, M.~Osadchy, L.~Ashton, M.~Foster, C.~J. Solomon and S.~J. Gibson, \emph{Analyst}, 2017, \textbf{142}, 4067--4074\relax
\mciteBstWouldAddEndPuncttrue
\mciteSetBstMidEndSepPunct{\mcitedefaultmidpunct}
{\mcitedefaultendpunct}{\mcitedefaultseppunct}\relax
\EndOfBibitem
\bibitem[Oviedo \emph{et~al.}(2019)Oviedo, Ren, Sun, Settens, Liu, Hartono, Ramasamy, DeCost, Tian, Romano,\emph{et~al.}]{oviedo2019fast}
F.~Oviedo, Z.~Ren, S.~Sun, C.~Settens, Z.~Liu, N.~T.~P. Hartono, S.~Ramasamy, B.~L. DeCost, S.~I. Tian, G.~Romano \emph{et~al.}, \emph{npj Computational Materials}, 2019, \textbf{5}, 60\relax
\mciteBstWouldAddEndPuncttrue
\mciteSetBstMidEndSepPunct{\mcitedefaultmidpunct}
{\mcitedefaultendpunct}{\mcitedefaultseppunct}\relax
\EndOfBibitem
\bibitem[Szymanski \emph{et~al.}(2023)Szymanski, Nevatia, Bartel, Zeng, and Ceder]{szymanski2023autonomous}
N.~J. Szymanski, P.~Nevatia, C.~J. Bartel, Y.~Zeng and G.~Ceder, \emph{Nature Communications}, 2023, \textbf{14}, 6956\relax
\mciteBstWouldAddEndPuncttrue
\mciteSetBstMidEndSepPunct{\mcitedefaultmidpunct}
{\mcitedefaultendpunct}{\mcitedefaultseppunct}\relax
\EndOfBibitem
\bibitem[Aykol \emph{et~al.}(2021)Aykol, Montoya, and Hummelshøj]{aykol2021rational}
M.~Aykol, J.~H. Montoya and J.~S. Hummelshøj, \emph{Journal of the American Chemical Society}, 2021, \textbf{143}, 9244--9259\relax
\mciteBstWouldAddEndPuncttrue
\mciteSetBstMidEndSepPunct{\mcitedefaultmidpunct}
{\mcitedefaultendpunct}{\mcitedefaultseppunct}\relax
\EndOfBibitem
\bibitem[Strieth-Kalthoff \emph{et~al.}(2023)Strieth-Kalthoff, Hao, Rathore, Derasp, Gaudin, Angello, Seifrid, Trushina, Guy, Liu, Tang, Mamada, Wang, Tsagaantsooj, Lavigne, Pollice, Wu, Hotta, Bodo, Li, Haddadnia, Wolos, Roszak, Ser, Bozal-Ginesta, Hickman, Vestfrid, Aguilar-Gránda, Klimareva, Sigerson, Hou, Gahler, Lach, Warzybok, Borodin, Rohrbach, Sanchez-Lengeling, Adachi, Grzybowski, Cronin, Hein, Burke, and Aspuru-Guzik]{strieth2023delocalized}
F.~Strieth-Kalthoff, H.~Hao, V.~Rathore, J.~Derasp, T.~Gaudin, N.~H. Angello, M.~Seifrid, E.~Trushina, M.~Guy, J.~Liu, X.~Tang, M.~Mamada, W.~Wang, T.~Tsagaantsooj, C.~Lavigne, R.~Pollice, T.~C. Wu, K.~Hotta, L.~Bodo, S.~Li, M.~Haddadnia, A.~Wolos, R.~Roszak, C.-T. Ser, C.~Bozal-Ginesta, R.~J. Hickman, J.~Vestfrid, A.~Aguilar-Gránda, E.~L. Klimareva, R.~C. Sigerson, W.~Hou, D.~Gahler, S.~Lach, A.~Warzybok, O.~Borodin, S.~Rohrbach, B.~Sanchez-Lengeling, C.~Adachi, B.~A. Grzybowski, L.~Cronin, J.~E. Hein, M.~D. Burke and A.~Aspuru-Guzik, \emph{ChemRxiv}, 2023\relax
\mciteBstWouldAddEndPuncttrue
\mciteSetBstMidEndSepPunct{\mcitedefaultmidpunct}
{\mcitedefaultendpunct}{\mcitedefaultseppunct}\relax
\EndOfBibitem
\bibitem[Granda \emph{et~al.}(2018)Granda, Donina, Dragone, Long, and Cronin]{granda2018controlling}
J.~M. Granda, L.~Donina, V.~Dragone, D.-L. Long and L.~Cronin, \emph{Nature}, 2018, \textbf{559}, 377--381\relax
\mciteBstWouldAddEndPuncttrue
\mciteSetBstMidEndSepPunct{\mcitedefaultmidpunct}
{\mcitedefaultendpunct}{\mcitedefaultseppunct}\relax
\EndOfBibitem
\bibitem[Xie \emph{et~al.}(2023)Xie, Sattari, Zhang, and Lin]{xie2022toward}
Y.~Xie, K.~Sattari, C.~Zhang and J.~Lin, \emph{Progress in Materials Science}, 2023, \textbf{132}, 101043\relax
\mciteBstWouldAddEndPuncttrue
\mciteSetBstMidEndSepPunct{\mcitedefaultmidpunct}
{\mcitedefaultendpunct}{\mcitedefaultseppunct}\relax
\EndOfBibitem
\bibitem[Abolhasani and Kumacheva(2023)]{abolhasani2023rise}
M.~Abolhasani and E.~Kumacheva, \emph{Nature Synthesis}, 2023, \textbf{2}, 483--492\relax
\mciteBstWouldAddEndPuncttrue
\mciteSetBstMidEndSepPunct{\mcitedefaultmidpunct}
{\mcitedefaultendpunct}{\mcitedefaultseppunct}\relax
\EndOfBibitem
\bibitem[Martin \emph{et~al.}(2023)Martin, Radivojevic, Zucker, Bouchard, Sustarich, Peisert, Arnold, Hillson, Babnigg, Marti,\emph{et~al.}]{martin2023perspectives}
H.~G. Martin, T.~Radivojevic, J.~Zucker, K.~Bouchard, J.~Sustarich, S.~Peisert, D.~Arnold, N.~Hillson, G.~Babnigg, J.~M. Marti \emph{et~al.}, \emph{Current Opinion in Biotechnology}, 2023, \textbf{79}, 102881\relax
\mciteBstWouldAddEndPuncttrue
\mciteSetBstMidEndSepPunct{\mcitedefaultmidpunct}
{\mcitedefaultendpunct}{\mcitedefaultseppunct}\relax
\EndOfBibitem
\bibitem[Xie \emph{et~al.}(2023)Xie, Sattari, Zhang, and Lin]{xie2023toward}
Y.~Xie, K.~Sattari, C.~Zhang and J.~Lin, \emph{Progress in Materials Science}, 2023, \textbf{132}, 101043\relax
\mciteBstWouldAddEndPuncttrue
\mciteSetBstMidEndSepPunct{\mcitedefaultmidpunct}
{\mcitedefaultendpunct}{\mcitedefaultseppunct}\relax
\EndOfBibitem
\bibitem[Raccuglia \emph{et~al.}(2016)Raccuglia, Elbert, Adler, Falk, Wenny, Mollo, Zeller, Friedler, Schrier, and Norquist]{raccuglia2016machine}
P.~Raccuglia, K.~C. Elbert, P.~D. Adler, C.~Falk, M.~B. Wenny, A.~Mollo, M.~Zeller, S.~A. Friedler, J.~Schrier and A.~J. Norquist, \emph{Nature}, 2016, \textbf{533}, 73--76\relax
\mciteBstWouldAddEndPuncttrue
\mciteSetBstMidEndSepPunct{\mcitedefaultmidpunct}
{\mcitedefaultendpunct}{\mcitedefaultseppunct}\relax
\EndOfBibitem
\bibitem[Zimmerman \emph{et~al.}(2014)Zimmerman, Grabowski, Domagalski, MacLean, Chruszcz, and Minor]{zimmerman2014data}
M.~D. Zimmerman, M.~Grabowski, M.~J. Domagalski, E.~M. MacLean, M.~Chruszcz and W.~Minor, \emph{Structural Genomics and Drug Discovery: Methods and Protocols}, 2014,  1--25\relax
\mciteBstWouldAddEndPuncttrue
\mciteSetBstMidEndSepPunct{\mcitedefaultmidpunct}
{\mcitedefaultendpunct}{\mcitedefaultseppunct}\relax
\EndOfBibitem
\bibitem[Talirz \emph{et~al.}(2020)Talirz, Kumbhar, Passaro, Yakutovich, Granata, Gargiulo, Borelli, Uhrin, Huber, Zoupanos,\emph{et~al.}]{talirz2020materials}
L.~Talirz, S.~Kumbhar, E.~Passaro, A.~V. Yakutovich, V.~Granata, F.~Gargiulo, M.~Borelli, M.~Uhrin, S.~P. Huber, S.~Zoupanos \emph{et~al.}, \emph{Scientific data}, 2020, \textbf{7}, 299\relax
\mciteBstWouldAddEndPuncttrue
\mciteSetBstMidEndSepPunct{\mcitedefaultmidpunct}
{\mcitedefaultendpunct}{\mcitedefaultseppunct}\relax
\EndOfBibitem
\bibitem[Jain \emph{et~al.}(2015)Jain, Ong, Chen, Medasani, Qu, Kocher, Brafman, Petretto, Rignanese, Hautier,\emph{et~al.}]{jain2015fireworks}
A.~Jain, S.~P. Ong, W.~Chen, B.~Medasani, X.~Qu, M.~Kocher, M.~Brafman, G.~Petretto, G.-M. Rignanese, G.~Hautier \emph{et~al.}, \emph{Concurrency and Computation: Practice and Experience}, 2015, \textbf{27}, 5037--5059\relax
\mciteBstWouldAddEndPuncttrue
\mciteSetBstMidEndSepPunct{\mcitedefaultmidpunct}
{\mcitedefaultendpunct}{\mcitedefaultseppunct}\relax
\EndOfBibitem
\bibitem[Himanen \emph{et~al.}(2019)Himanen, Geurts, Foster, and Rinke]{himanen2019data}
L.~Himanen, A.~Geurts, A.~S. Foster and P.~Rinke, \emph{Advanced Science}, 2019, \textbf{6}, 1900808\relax
\mciteBstWouldAddEndPuncttrue
\mciteSetBstMidEndSepPunct{\mcitedefaultmidpunct}
{\mcitedefaultendpunct}{\mcitedefaultseppunct}\relax
\EndOfBibitem
\bibitem[Curtarolo \emph{et~al.}(2012)Curtarolo, Setyawan, Hart, Jahnatek, Chepulskii, Taylor, Wang, Xue, Yang, Levy,\emph{et~al.}]{curtarolo2012aflow}
S.~Curtarolo, W.~Setyawan, G.~L. Hart, M.~Jahnatek, R.~V. Chepulskii, R.~H. Taylor, S.~Wang, J.~Xue, K.~Yang, O.~Levy \emph{et~al.}, \emph{Computational Materials Science}, 2012, \textbf{58}, 218--226\relax
\mciteBstWouldAddEndPuncttrue
\mciteSetBstMidEndSepPunct{\mcitedefaultmidpunct}
{\mcitedefaultendpunct}{\mcitedefaultseppunct}\relax
\EndOfBibitem
\bibitem[Mathew \emph{et~al.}(2017)Mathew, Montoya, Faghaninia, Dwarakanath, Aykol, Tang, Chu, Smidt, Bocklund, Horton,\emph{et~al.}]{mathew2017atomate}
K.~Mathew, J.~H. Montoya, A.~Faghaninia, S.~Dwarakanath, M.~Aykol, H.~Tang, I.-h. Chu, T.~Smidt, B.~Bocklund, M.~Horton \emph{et~al.}, \emph{Computational Materials Science}, 2017, \textbf{139}, 140--152\relax
\mciteBstWouldAddEndPuncttrue
\mciteSetBstMidEndSepPunct{\mcitedefaultmidpunct}
{\mcitedefaultendpunct}{\mcitedefaultseppunct}\relax
\EndOfBibitem
\bibitem[Huber \emph{et~al.}(2021)Huber, Bosoni, Bercx, Br{\"o}der, Degomme, Dikan, Eimre, Flage-Larsen, Garcia, Genovese,\emph{et~al.}]{huber2021common}
S.~P. Huber, E.~Bosoni, M.~Bercx, J.~Br{\"o}der, A.~Degomme, V.~Dikan, K.~Eimre, E.~Flage-Larsen, A.~Garcia, L.~Genovese \emph{et~al.}, \emph{npj Computational Materials}, 2021, \textbf{7}, 136\relax
\mciteBstWouldAddEndPuncttrue
\mciteSetBstMidEndSepPunct{\mcitedefaultmidpunct}
{\mcitedefaultendpunct}{\mcitedefaultseppunct}\relax
\EndOfBibitem
\bibitem[Sim \emph{et~al.}(2023)Sim, Vakili, Strieth-Kalthoff, Hao, Hickman, Miret, Pablo-Garc{\'\i}a, and Aspuru-Guzik]{sim2023chemos}
M.~Sim, M.~G. Vakili, F.~Strieth-Kalthoff, H.~Hao, R.~Hickman, S.~Miret, S.~Pablo-Garc{\'\i}a and A.~Aspuru-Guzik, \emph{ChemRxiv}, 2023\relax
\mciteBstWouldAddEndPuncttrue
\mciteSetBstMidEndSepPunct{\mcitedefaultmidpunct}
{\mcitedefaultendpunct}{\mcitedefaultseppunct}\relax
\EndOfBibitem
\bibitem[Juchli(2022)]{Juchli2022}
D.~Juchli, in \emph{SiLA 2: The Next Generation Lab Automation Standard}, ed. S.~Beutel and F.~Lenk, Springer International Publishing, Cham, 2022, pp. 147--174\relax
\mciteBstWouldAddEndPuncttrue
\mciteSetBstMidEndSepPunct{\mcitedefaultmidpunct}
{\mcitedefaultendpunct}{\mcitedefaultseppunct}\relax
\EndOfBibitem
\bibitem[Rahmanian \emph{et~al.}(2022)Rahmanian, Flowers, Guevarra, Richter, Fichtner, Donnely, Gregoire, and Stein]{rahmanian2022enabling}
F.~Rahmanian, J.~Flowers, D.~Guevarra, M.~Richter, M.~Fichtner, P.~Donnely, J.~M. Gregoire and H.~S. Stein, \emph{Advanced Materials Interfaces}, 2022, \textbf{9}, 2101987\relax
\mciteBstWouldAddEndPuncttrue
\mciteSetBstMidEndSepPunct{\mcitedefaultmidpunct}
{\mcitedefaultendpunct}{\mcitedefaultseppunct}\relax
\EndOfBibitem
\bibitem[Guevarra \emph{et~al.}(2023)Guevarra, Kan, Lai, Jones, Zhou, Donnelly, Richter, Stein, and Gregoire]{guevarra2023orchestrating}
D.~Guevarra, K.~Kan, Y.~Lai, R.~J. Jones, L.~Zhou, P.~Donnelly, M.~Richter, H.~S. Stein and J.~M. Gregoire, \emph{Digital Discovery}, 2023, \textbf{2}, 1806--1812\relax
\mciteBstWouldAddEndPuncttrue
\mciteSetBstMidEndSepPunct{\mcitedefaultmidpunct}
{\mcitedefaultendpunct}{\mcitedefaultseppunct}\relax
\EndOfBibitem
\bibitem[Allan \emph{et~al.}(2019)Allan, Caswell, Campbell, and Rakitin]{allan2019bluesky}
D.~Allan, T.~Caswell, S.~Campbell and M.~Rakitin, \emph{Synchrotron Radiation News}, 2019, \textbf{32}, 19--22\relax
\mciteBstWouldAddEndPuncttrue
\mciteSetBstMidEndSepPunct{\mcitedefaultmidpunct}
{\mcitedefaultendpunct}{\mcitedefaultseppunct}\relax
\EndOfBibitem
\bibitem[Fakhruldeen \emph{et~al.}(2022)Fakhruldeen, Pizzuto, Glowacki, and Cooper]{fakhruldeen2022archemist}
H.~Fakhruldeen, G.~Pizzuto, J.~Glowacki and A.~I. Cooper, 2022 International Conference on Robotics and Automation (ICRA), 2022, pp. 6013--6019\relax
\mciteBstWouldAddEndPuncttrue
\mciteSetBstMidEndSepPunct{\mcitedefaultmidpunct}
{\mcitedefaultendpunct}{\mcitedefaultseppunct}\relax
\EndOfBibitem
\bibitem[Pendleton \emph{et~al.}(2019)Pendleton, Cattabriga, Li, Najeeb, Friedler, Norquist, Chan, and Schrier]{pendleton2019experiment}
I.~M. Pendleton, G.~Cattabriga, Z.~Li, M.~A. Najeeb, S.~A. Friedler, A.~J. Norquist, E.~M. Chan and J.~Schrier, \emph{MRS Communications}, 2019, \textbf{9}, 846--859\relax
\mciteBstWouldAddEndPuncttrue
\mciteSetBstMidEndSepPunct{\mcitedefaultmidpunct}
{\mcitedefaultendpunct}{\mcitedefaultseppunct}\relax
\EndOfBibitem
\bibitem[Vescovi \emph{et~al.}(2023)Vescovi, Ginsburg, Hippe, Ozgulbas, Stone, Stroka, Butler, Blaiszik, Brettin, Chard,\emph{et~al.}]{vescovi2023towards}
R.~Vescovi, T.~Ginsburg, K.~Hippe, D.~Ozgulbas, C.~Stone, A.~Stroka, R.~Butler, B.~Blaiszik, T.~Brettin, K.~Chard \emph{et~al.}, \emph{Digital Discovery}, 2023, \textbf{2}, 1980--1998\relax
\mciteBstWouldAddEndPuncttrue
\mciteSetBstMidEndSepPunct{\mcitedefaultmidpunct}
{\mcitedefaultendpunct}{\mcitedefaultseppunct}\relax
\EndOfBibitem
\bibitem[Zhou \emph{et~al.}(2024)Zhou, Luo, Chen, Zhu, Jiang, Zhang, Shang, and Jiang]{zhou2024multi}
J.~Zhou, M.~Luo, L.~Chen, Q.~Zhu, S.~Jiang, F.~Zhang, W.~Shang and J.~Jiang, \emph{ChemRxiv}, 2024\relax
\mciteBstWouldAddEndPuncttrue
\mciteSetBstMidEndSepPunct{\mcitedefaultmidpunct}
{\mcitedefaultendpunct}{\mcitedefaultseppunct}\relax
\EndOfBibitem
\bibitem[{The Pydantic development team}(2024)]{pydantic2024}
{The Pydantic development team}, \emph{Pydantic}, \url{https://github.com/pydantic/pydantic}, 2024, Accessed: 2024-02-05\relax
\mciteBstWouldAddEndPuncttrue
\mciteSetBstMidEndSepPunct{\mcitedefaultmidpunct}
{\mcitedefaultendpunct}{\mcitedefaultseppunct}\relax
\EndOfBibitem
\bibitem[Bertolotti and Hu(2017)]{swales1999open}
I.~C. Bertolotti and T.~Hu, \emph{Embedded Software Development}, 2017, \textbf{29}, 343--374\relax
\mciteBstWouldAddEndPuncttrue
\mciteSetBstMidEndSepPunct{\mcitedefaultmidpunct}
{\mcitedefaultendpunct}{\mcitedefaultseppunct}\relax
\EndOfBibitem
\bibitem[Pezoa \emph{et~al.}(2016)Pezoa, Reutter, Suarez, Ugarte, and Vrgo{\v{c}}]{pezoa2016foundations}
F.~Pezoa, J.~L. Reutter, F.~Suarez, M.~Ugarte and D.~Vrgo{\v{c}}, Proceedings of the 25th International Conference on World Wide Web, 2016, pp. 263--273\relax
\mciteBstWouldAddEndPuncttrue
\mciteSetBstMidEndSepPunct{\mcitedefaultmidpunct}
{\mcitedefaultendpunct}{\mcitedefaultseppunct}\relax
\EndOfBibitem
\bibitem[Jeffay \emph{et~al.}(1991)Jeffay, Stanat, and Martel]{jeffay1991non}
K.~Jeffay, D.~F. Stanat and C.~U. Martel, IEEE real-time systems symposium, 1991, pp. 129--139\relax
\mciteBstWouldAddEndPuncttrue
\mciteSetBstMidEndSepPunct{\mcitedefaultmidpunct}
{\mcitedefaultendpunct}{\mcitedefaultseppunct}\relax
\EndOfBibitem
\bibitem[Mi \emph{et~al.}(2017)Mi, Qian, Zhang, and Wang]{mi2017empirical}
X.~Mi, F.~Qian, Y.~Zhang and X.~Wang, Proceedings of the 2017 Internet Measurement Conference, 2017, pp. 398--404\relax
\mciteBstWouldAddEndPuncttrue
\mciteSetBstMidEndSepPunct{\mcitedefaultmidpunct}
{\mcitedefaultendpunct}{\mcitedefaultseppunct}\relax
\EndOfBibitem
\bibitem[Srinivasan(1995)]{srinivasan1995rfc1831}
R.~Srinivasan, \emph{RFC1831: RPC: Remote Procedure Call Protocol Specification Version 2}, 1995\relax
\mciteBstWouldAddEndPuncttrue
\mciteSetBstMidEndSepPunct{\mcitedefaultmidpunct}
{\mcitedefaultendpunct}{\mcitedefaultseppunct}\relax
\EndOfBibitem
\bibitem[H{\"a}se \emph{et~al.}(2018)H{\"a}se, Roch, and Aspuru-Guzik]{hase2018chimera}
F.~H{\"a}se, L.~M. Roch and A.~Aspuru-Guzik, \emph{Chemical science}, 2018, \textbf{9}, 7642--7655\relax
\mciteBstWouldAddEndPuncttrue
\mciteSetBstMidEndSepPunct{\mcitedefaultmidpunct}
{\mcitedefaultendpunct}{\mcitedefaultseppunct}\relax
\EndOfBibitem
\bibitem[Hickman \emph{et~al.}(2023)Hickman, Sim, Pablo-Garc{\'\i}a, Woolhouse, Hao, Bao, Bannigan, Allen, Aldeghi, and Aspuru-Guzik]{hickman_atlas_2023}
R.~Hickman, M.~Sim, S.~Pablo-Garc{\'\i}a, I.~Woolhouse, H.~Hao, Z.~Bao, P.~Bannigan, C.~Allen, M.~Aldeghi and A.~Aspuru-Guzik, \emph{Atlas: {A} {Brain} for {Self}-driving {Laboratories}}, 2023\relax
\mciteBstWouldAddEndPuncttrue
\mciteSetBstMidEndSepPunct{\mcitedefaultmidpunct}
{\mcitedefaultendpunct}{\mcitedefaultseppunct}\relax
\EndOfBibitem
\end{mcitethebibliography}
\bibliographystyle{rsc} 

\end{document}